\documentclass[fleqn,usenatbib]{mnras}

\usepackage{graphicx}	
\usepackage{amsmath}	
\usepackage{amssymb}	
\usepackage{multicol}        
\usepackage{bm}		
\usepackage{pdflscape}	
\usepackage{comment}

\newcommand{\kms}{{\hbox{km s$^{-1}$}}} 

\newcommand{\ebv}{\hbox{$E(B-V)$}}
\newcommand{\Msol}{M$_\odot$}

\usepackage[T1]{fontenc}
\usepackage{ae,aecompl}

\usepackage{newtxtext,newtxmath}

\title[Correlations of star clusters with GMCs in M51]{The Spatial Relation between Young Star Clusters and Molecular Clouds in M~51 with LEGUS}

\author[K. Grasha et al.]{
K. Grasha,$^{1,2,3}$\thanks{E-mail: kathryn.grasha@anu.edu.au}
D. Calzetti,$^{1}$
A. Adamo,$^{4}$
R.C. Kennicutt,$^{5}$
B.G. Elmegreen,$^{6}$
\newauthor
M. Messa,$^{4}$
D.A. Dale,$^{7}$
K. Fedorenko,$^{8}$
S. Mahadevan,$^{8,9}$
E.K. Grebel,$^{10}$
\newauthor
M. Fumagalli,$^{11}$
H. Kim,$^{12}$
C.L. Dobbs,$^{13}$
D.A. Gouliermis,$^{14,15}$
G. Ashworth,$^{11}$
\newauthor
J.S. Gallagher III,$^{16}$
L.J. Smith,$^{17}$
M. Tosi,$^{18}$
B.C. Whitmore,$^{19}$
E. Schinnerer,$^{15}$
\newauthor
D. Colombo,$^{20}$
A. Hughes,$^{21}$
A.K. Leroy,$^{22}$
S.E. Meidt$^{15}$
\\
$^{1}$Astronomy Department, University of Massachusetts, Amherst, MA 01003, USA; kathryn.grasha@anu.edu.au\\
$^{2}$Research School of Astronomy and Astrophysics, Australian National University, Canberra, ACT 2611, Australia\\
$^{3}$ARC Centre of Excellence for All Sky Astrophysics in 3 Dimensions (ASTRO 3D), Australia\\
$^{4}$Department of Astronomy, The Oskar Klein Centre, Stockholm University, SE-106 91 Stockholm, Sweden\\
$^{5}$Institute of Astronomy, University of Cambridge, Cambridge CB3 0HA, UK\\
$^{6}$IBM Research Division, T.J. Watson Research Center, Yorktown Hts., NY 10598, USA\\
$^{7}$Department of Physics and Astronomy, University of Wyoming, Laramie, WY 82071, USA\\
$^{8}$College of Information and Computer Sciences, University of Massachusetts, Amherst, MA 01003, USA\\
$^{9}$Stanford Research Institute, Menlo Park, CA 94025, USA\\
$^{10}$Astronomisches Rechen-Institut, Zentrum f\"ur Astronomie der Universit\"at Heidelberg, M\"onchhofstr.\ 12--14, 69120 Heidelberg, Germany\\
$^{11}$Institute for Computational Cosmology and Centre for Extragalactic Astronomy, Durham University, Durham DH1 3LE, UK\\
$^{12}$Gemini Observatory, Casilla 603, La Serena, Chile\\
$^{13}$School of Physics, University of Exeter, Stocker Road, Exeter EX4 4QL, UK\\
$^{14}$Zentrum f\"ur Astronomie der Universit\"at Heidelberg, Institut f\"ur Theoretische Astrophysik, Albert-Ueberle-Str.\,2, 69120 Heidelberg, Germany\\
$^{15}$Max Planck Institute for Astronomy, K\"{o}nigstuhl\,17, 69117 Heidelberg, Germany\\
$^{16}$Department of Astronomy, University of Wisconsin--Madison, Madison, WI 53706, USA\\
$^{17}$European Space Agency/Space Telescope Science Institute, Baltimore, MD 21218, USA\\
$^{18}$INAF -- Osservatorio di Astrofisica e Scienza dello Spazio, I-40127 Bologna, Italy\\
$^{19}$Space Telescope Science Institute, Baltimore, MD 21218, USA\\
$^{20}$MPI for Radioastronomy, Auf dem Hgel, D-53010 Bonn, Germany\\
$^{21}$IRAP, 9 av. du Colonel Roche, BP 44346, F-31028 Toulouse cedex 4, France\\
$^{22}$Department of Astronomy, The Ohio State University, 140 West 18th Avenue, Columbus, OH 43210, USA
}

\pubyear{2018}

\begin{document}
\label{firstpage}
\pagerange{\pageref{firstpage}--\pageref{lastpage}}
\maketitle

\begin{abstract} 
We present a study correlating the spatial locations of young star clusters with those of molecular clouds in NGC~5194, in order to investigate the timescale over which clusters separate from their birth clouds. The star cluster catalogues are from the Legacy ExtraGalactic UV Survey (LEGUS) and the molecular clouds from the Plateau de Bure Interefrometer Arcsecond Whirpool Survey (PAWS). We find that younger star clusters are spatially closer to molecular clouds than older star clusters. The median ages for clusters associated with clouds is 4~Myr whereas it is 50~Myr for clusters that are sufficiently separated from a molecular cloud to be considered unassociated. After $\sim$6~Myr, the majority of the star clusters lose association with their molecular gas. Younger star clusters are also preferentially located in stellar spiral arms where they are hierarchically distributed in kpc-size regions for 50--100~Myr before dispersing. The youngest star clusters are more strongly clustered, yielding a two-point correlation function with $\alpha=-0.28\pm0.04$, than the GMCs ($\alpha=-0.09\pm0.03$) within the same PAWS field. However, the clustering strength of the most massive GMCs, supposedly the progenitors of the young clusters for a star formation efficiency of a few percent, is comparable ($\alpha=-0.35\pm0.05$) to that of the clusters. We find a galactocentric-dependence for the coherence of star formation, in which clusters located in the inner region of the galaxy reside in smaller star-forming complexes and display more homogeneous distributions than clusters further from the centre. This result suggests a correlation between the survival of a cluster complex and its environment.
\end{abstract}




\begingroup
\let\clearpage\relax
\endgroup
\newpage

\section{Introduction}\label{sec:intro}
A central tenet of our understanding of star formation is that stars form in stellar aggregates \citep[e.g.,][]{lada03,portegieszwart10} that we call `clustered star formation'. In such a distribution, individual components of a population are more likely to appear near other members than if they were randomly distributed. Observations have directly measured the correlation in the spatial distribution between young stars, stellar clusters, and associations \citep[e.g.,][]{gomez93, zhang01, oey04, delafuentemarcos09, karampelas09, bianchi12, gouliermis15, gouliermis17, gouliermis18, grasha15, grasha17a, grasha17b, sun17a, sun17b, rodriguez18}, providing overwhelming evidence for the coherence of star formation over multiple scales and across most galactic environments. 

This hierarchical distribution of star formation is understood as a consequence of dense gas concentrated primarily within giant molecular clouds (GMCs) that undergo fragmentation \citep{carlberg90, mclaughlin96}, under the influence of both gravitational collapse \citep{devega96, kuznetsova18} and turbulence \citep[e.g.,][]{elmegreenscalo04, federrath09, girichidis12, hopkinsetal13, guszejnov16}, and other feedback processes that act to suppress star formation \citep[e.g.,][]{krumholz14}. The hierarchical fractal structure of the interstellar medium (ISM) results in a power-law distribution of the ISM components from which we can measure the fractal dimension of the distribution \citep[e.g.,][]{elmegreenfalgarone96, romanduval10}. Within this framework, young star clusters should inherit their clustered distribution from the ISM structure from which they are born; observations corroborate the hierarchy of gas \citep[e.g.,][]{elmegreenfalgarone96, stutzki98, sanchez10, miville10, elia18} as well as protostellar cores \citep{johnstone00, johnstone01, enoch06, stanke06, young06} and young stellar objects \citep{bressert10, sun18}. 

Star clusters, identifiable in galaxies up to distances of $\sim$100~Mpc, are born at the densest peaks of the hierarchy \citep{elmegreen08} and provide a sensitive and direct observational signature of the star formation process. The hierarchical distribution of star clusters is slowly lost with time, either due to random motions that remove the fractal imprint, becoming more uniformly distributed with age, or due to the merging of sub-clusters into larger clusters \citep[e.g.,][]{gieles08, bastian09, davidge11}. Despite their exceptional tracers of recent star formation, it is not well understood to what extent the galactic environment, both locally and globally, influences the evolution and survival of star clusters or the complexes from which they are born. These outstanding issues need to be properly addressed in order to accurately characterize the connection between star formation occurring at the scales of individual stellar aggregates, with the global scaling relation between star formation and gas reservoirs of entire galaxies \citep{kennicutt98, kennicutt12}.

M~51 (NGC~5194 and NGC~5195) is part of the Legacy ExtraGalactic UV Survey\footnote{https://archive.stsci.edu/prepds/legus/} \citep[LEGUS, HST GO--13364;][]{calzetti15a}, a Cycle 21 HST treasury program of 50 nearby ($\sim$3.5--16~Mpc) galaxies in five UV and optical bands (NUV, U, B, V, and I) with the goal of investigating the connection between local star forming regions --- as traced with young stellar clusters --- and global star formation within the nearby universe. Results with the LEGUS datasets so far include support for a hierarchical star formation process \citep{elmegreen14, gouliermis17, grasha15, grasha17a} with the age and size distribution of the hierarchies driven by turbulence \citep{gouliermis15, grasha17b}; investigation of the effect on the evolution of galaxies from the radiative and mechanical feedback of star clusters \citep{calzetti15b}; tests for variations in the initial mass function of star clusters \citep{krumholz15a, ashworth17}; test spiral density wave theory \citep{shabani18}; and extinction maps using stellar catalogues to investigate variations in the dust--to--gas ratio with metallicity \citep{kahre18}. In this work, we perform a cross-correlation analysis between the star clusters from the LEGUS catalogue and the GMC catalogue from the PAWS project in M~51. Very little exists in the literature, with the only other analysis having been performed in the flocculent spiral NGG~7793 \citep{grasha18}.

The cluster catalogue of M~51 covers a much larger portion of the galaxy than the catalogue of NGC~7793. Thus, M~51 is an excellent benchmark to investigate the clustering nature of star clusters as a function of galactocentric distance in a grand-design spiral, in addition to the connection of the young stellar clusters with maps of molecular clouds, both investigated in this paper. The former has been the main topic of two previous papers \citep{grasha15, grasha17a}, but not yet performed in such a cluster-rich galaxy to explore trends at sub-galactic scales. 

The paper is organized as follows: the galaxy selection and reduction process are described in Section \ref{sec:sample}. The cluster selection and identification process are described in Section~\ref{sec:clusterselection}. The results are described in Section \ref{sec:results}, where Section \ref{sec:gas} briefly introduces the molecular gas data and the two-point correlation function is described in Section \ref{sec:2pcf}. We discuss our results concerning hierarchy of the stellar clusters and the connection to the gas reservoirs in Section \ref{sec:discussion}. We summarize the findings of this study in Section \ref{sec:summary}.

\section{Sample Selection and Data Reduction}\label{sec:sample}
In this paper, we select M~51 (NGC~5194 with its smaller interacting companion, NGC~5195, collectively known as the Whirlpool galaxy) from the LEGUS survey due to the large number of cluster candidates over a large observed field with strong spiral features.

NGC~5194 is relatively large in size (angular size of $\sim$11' $\times$ 7' and stellar mass $M_{\star} = 2.4\times 10^{10}$~M$_{\odot}$), and combined with a star formation rate SFR(UV) = 3 M$_{\odot}$~yr$^{-1}$, this system provides a large and robust number of young clusters. NGC~5194 has a relatively low inclination \citep[22$^{\circ}$;][]{colombo14b}, mitigating projection effects, and a distance of 7.66~Mpc \citep{tonry01}, sufficiently nearby to lessen confusion between stellar systems and individual stars. 

\begin{figure*}
\includegraphics[scale=.55]{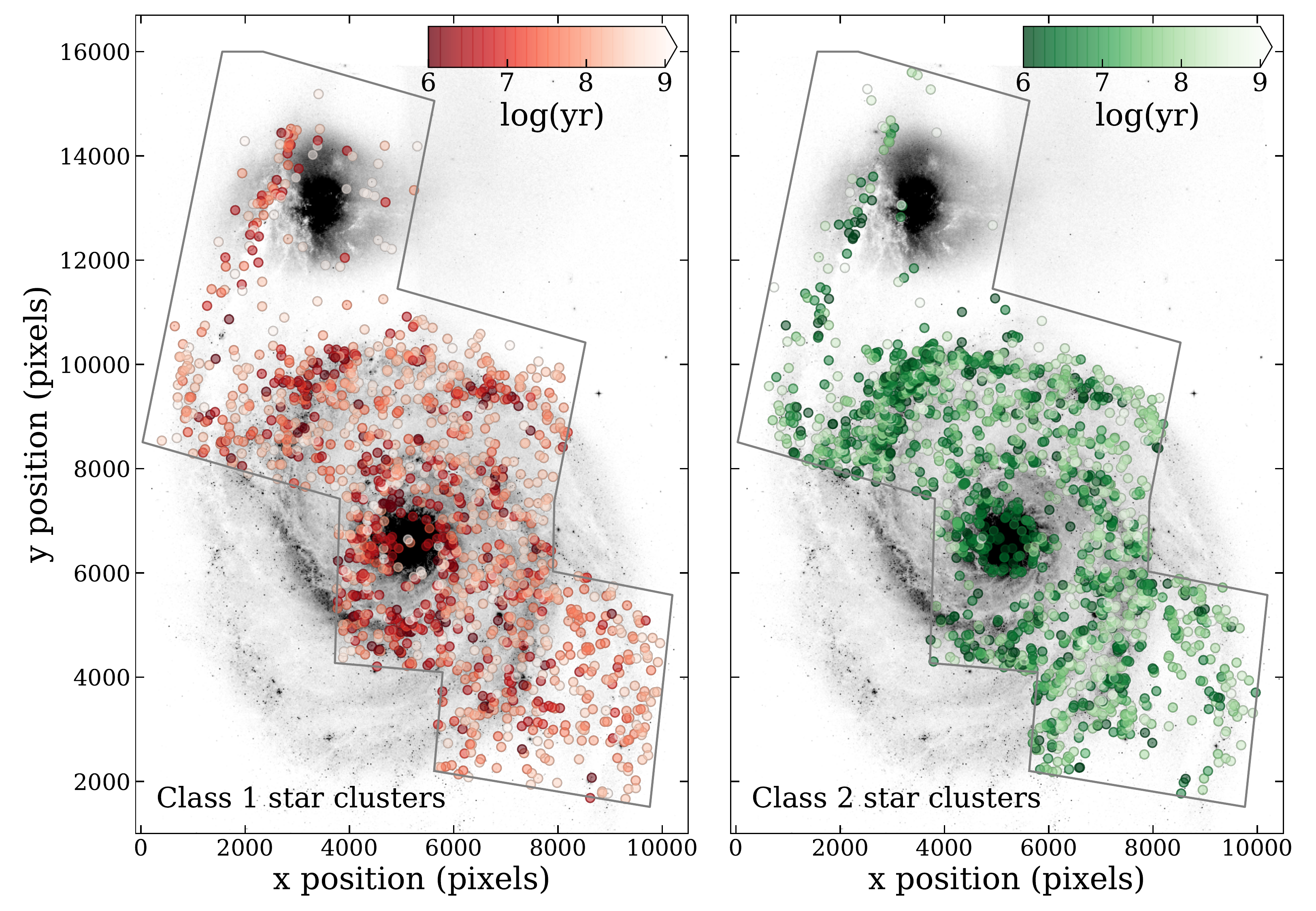}
\caption{
The V-band ACS image of NGC~5194 overplotted with the location of the class 1 (left; red) and class 2 (right; green) star clusters. The star clusters are colored according to their ages with dark colors corresponding to younger ages and all clusters older than 1 Gyr are shown as white. The black outline shows the UVIS footprint. Cluster classifications are based off of morphology and our morphological cluster classification is a good approximation to also a dynamical classification \citep{adamo17, grasha17a, ryon17}. The typical ages of class 1 clusters are older and show median ages of 90~Myr than the median age of 20~Myr for the class 2 clusters. The class 1 sources are much more uniformly distributed while the class 2 clusters are predominantly tracing the spiral arms and the center region. The youngest clusters of both classes are mostly concentrated along the spiral arms. 
\label{fig:5194}}
\end{figure*}

The LEGUS dataset of NGC~5194 provides 5 pointings in the NUV (F275W) and U (F336W) broad-bands, observed with WFC3/UVIS. The new UV/U data consist of 4 pointings from LEGUS, covering the arms and outskirts of the galaxy with a deeper exposure covering the central nucleus of the galaxy (GO--13340; PI: S. Van Dyk). Archival B (F435W), V(F555W), and I (F814W) images with ACS/WFC (GO--10452; PI: S. Beckwith) cover the entire NGC~5194+NGC~5195 system with 6 pointings, re-reduced using the same pipeline with the UV and U WFC3/UVIS images from the LEGUS project. The UVIS data only cover a portion of the galaxy (see Figure~\ref{fig:5194}), and as a result, our clusters are only located within the footprint of the UVIS data as these blue wavelengths are necessary in order to break the age/dust degeneracy and accurately derive the ages of the young stellar clusters \citep[e.g.,][]{anders04, chandar10b, adamo17}. 

Reduced science frames are drizzled to a common scale resolution, corresponding to the native WFC3 pixel size (0.0396 arcsec/pixel). The frames have all been aligned and rotated with North up. Detailed descriptions of the standard data reduction of the LEGUS datasets are available in \citet{calzetti15a}. 
%

\section{Creating the Visually-Identified Star Cluster catalogues}\label{sec:clusterselection}
A detailed description of the cluster selection, identification, photometry, and SED fitting procedures for the LEGUS galaxies is presented in \citet{adamo17}. The production of the cluster catalogue and completeness tests applied to NGC~5194 are detailed in \citet{messa18a} and the completeness limit in the five broad bands and across the various environmental sub-regions are further detailed in \citet{messa18b}. We summarize here briefly the aspects of \citeauthor{adamo17} and \citeauthor{messa18a} that are important for the current paper. 

Procuring cluster catalogues is a multi-step process: extracting the cluster candidates (Section~\ref{sec:SCselection}) through an automatic catalogue and then performing visual inspection of this catalogue to remove contaminants (Section~\ref{sec:SCidentification}). Completing these steps results in our final robust cluster catalogue.

\subsection{Star Cluster Selection}\label{sec:SCselection}
The {\it automated catalogue} of star cluster candidates is extracted from the V-band image with source extractor \citep[SExtractor;][]{bertin96}. For NGC~5194, the SExtractor input parameters are optimized to extract sources with at least a 10$\sigma$ detection in a minimum of 10 contiguous pixels. This automatic procedure returns the positions of candidate clusters within the image and the concentration index (CI; magnitude difference of each source within an aperture of 1 pixel compared to 3 pixels) of each source. The CI is related to the size of stellar systems \citep{ryon17} and can be used to differentiate between individual stars and stellar clusters; stars, in general, have narrower light profiles, and therefore, smaller CI values compared to star clusters. The CI value that separates stars from star clusters within each system and image is determined through an iterative inspection of the CI distribution from the output of the SExtractor parameters. The CI reference value used to distinguish between unresolved sources (stars) and resolved sources (candidate clusters) within NGC~5194 is 1.35 mag. There are 30176 sources satisfying the CI cut of 1.35~mag that make the automated catalogue. At the distance of NGC~5194, the pixel resolution is 1.47 parsec/pixel. 

All candidate clusters must satisfy two criteria in the above automated procedure: (1) the CI in the V-band must exceed the stellar CI peak value; and (2) be detected in at least two contiguous bands (the reference V band and either B or I band). Standard aperture photometry is performed for each cluster candidate using a fixed science aperture of 4 pixels in radius with a local sky annulus at 7 pixels (1 pixel wide) in all five filters. Aperture corrections to account for missing flux are based on isolated clusters \citep[see][]{messa18a} and calculated by subtracting the standard photometry in the fixed science aperture from the total photometry inside a 20 pixel radius with a 1 pixel sky annulus. Corrections for foreground Galactic extinction \citep{schlafly11} are applied to the photometry. 

Following this step, all cluster candidates detected in at least four bands with photometric error $\leq$0.3~mag undergo spectral energy distribution (SED) fitting procedures to procure the age, mass, and color excess \ebv\ of each source. The SED fitting analysis is performed with deterministic Yggdrasil single stellar population (SSP) models \citep{zackrisson11}. The Yggdrasil spectral synthesis code is based on Starburst99 \citep{leitherer99} stellar population spectra with the photoionized nebula predicted by Cloudy \citep{ferland98, ferland13}. All cluster catalogues for the LEGUS galaxies use a \citet{kroupa01} universal initial mass function (IMF; but see \citet{ashworth17} for a generalization to a variable IMF). The cluster properties in this paper are derived using Padova isochrones that include thermally pulsating asymptotic giant branch stars \citep{vazquez05} and the starburst attenuation curve \citep{calzetti00}. 

The fitting algorithm is based on a $\chi^2$ approached as described in \citet{adamo10} and the error analysis is described in \citet{adamo12}. The SED fitting method procedure produces average uncertainties of 0.1--0.2~dex in both the cluster ages and masses. As our cluster photometry is produced with average aperture corrections derived from observed clusters in each band, only the normalization of the cluster SED will be affected while the intrinsic shape of the SED as well as the intrinsic colors of the clusters are preserved. As a result, average aperture correction may affect the mass estimates of sources, however, any uncertainties in the mass introduced from aperture corrections will be within the typical 0.1--0.2~dex uncertainty measurements. 

We assume a fully sampled IMF for the derivation of our star cluster properties, however, star clusters below $\sim$3000~M$_{\odot}$ are subject to stochastic sampling of the IMF. The inclusion of the NUV stellar continuum in the SED fitting provides more photometric stability (by a factor of $\sim$3.5--4) relative to the H$\alpha$ emission \citep{calzetti10, andrews13}. This enables derivation of relatively accurate ages and masses of young ($<$10~Myr) star clusters down to $\sim$500~M$_{\odot}$ in our catalogues. The U--B colors provide an effective age indicator for the star clusters and the information provided at the NUV wavelengths discriminate between young and dusty clusters and old, dust-free clusters \citep{chandar10b}. \citet{calzetti15b} shows detailed SED fits for a range of star clusters in the LEGUS galaxy NGC~5253, demonstrating the fainter NUV flux and higher U--B colors in older star clusters compared to younger clusters.

\subsection{Visual Inspection \& Star Cluster Classification}\label{sec:SCidentification}
After the first step --- extraction of the clusters from an automatic catalogue and SED fitting --- all clusters with an absolute magnitude brighter than $-6$ mag in the V-band undergo visual inspection by a minimum of three independent classifiers within the LEGUS team to produce the final {\it visual catalogue}. The magnitude limit is defined by the detection limits of the LEGUS sample, which enables selecting down to a $\sim$1000~$M_{\odot}$, 6~Myr old cluster with \ebv\ = 0.25 \citep{calzetti15a}. 

A total of 10925 cluster candidates brighter than our magnitude cut off pass all the criteria in the extraction process (Section~\ref{sec:SCselection}). Of these, 2487 candidates were inspected visually as described above, and the remaining candidates were inspected through a bagged decision tree Machine Learning algorithm trained to reproduce the results of the human classifiers using the visually classified subset of class 1, 2, 3, and 4 as a training set \citep{g18}. 

The visual inspection step, either by humans or by a trained algorithm, is the only way to reliably remove non-cluster interlopers within the automatically extracted catalogue to ensure the robustness of the final cluster catalogue. While creating more conservative cuts in the automatic extraction parameters such as raising the CI value (i.e., selecting only the most extended sources) may exclude stellar contaminants, it is difficult to remove bad pixels, foreground stars, or background galaxies as those types of interlopers are not standardized in size or color. 

The human visual classification is performed using a custom interactive tool that utilizes DS9 with V-band and red-green-blue (RGB) composite images. Based on the morphology and color, each source gets classified under one of four classes as defined by the LEGUS collaboration \citep[see][]{adamo17}:

\begin{enumerate}
\item[] {\bf Class 1}: Symmetric and centrally compact star clusters. Usually uniform in color.
\item[] {\bf Class 2}: Compact, asymmetric star clusters with some degree of elongation. Usually uniform in color.
\item[] {\bf Class 3}: Compact associations that show multiple-peaked profiles on top of an underlying diffuse emission. 
\item[] {\bf Class 4}: Non-clusters, including but not limited to foreground stars, asterisms, background galaxies, saturated pixels, etc.
\end{enumerate}

The inspection of 2487 cluster candidates in the UVIS coverage (see Figure~\ref{fig:5194}) was performed by members of the LEGUS team, with a yield of 1226 (49\%) considered class 1, 2, or 3 clusters. The resulting agreement between classifiers is around 70\%--75\%. For large disagreements between the LEGUS classifiers (more than one class discrepancy between the classifiers), additional classifiers are involved, until a consistent classification for the candidate is achieved. This level of agreement between classifiers for LEGUS star clusters is the approximately the same accuracy as the visual classification of galaxies the CANDELS fields \citep{kartaltepe15} of $\sim$70\%. 

The visually classified catalogue was then used as a training set for the Machine Learning algorithm, which then proceeded to classify the remaining candidates in the automatic catalogue \citep{g18}. Our final cluster catalogue contains 3374 star clusters, and is publicly available\footnote{https://archive.stsci.edu/prepds/legus/dataproducts-public.html}. In the remaining of this paper we will concentrate on the properties of class 1 and 2 clusters, while class 3 (compact associations) and class 4 (contaminants) will no longer be considered.  There are a total of 2989 class 1 and 2 clusters in our final catalogue.

\subsection{Comparison of the Visual-Identified Clusters to the Machine Learning Clusters}\label{sec:compare}
Here we we briefly compare the properties of the visually identified star clusters to the ML-selected star clusters to demonstrate that the ML star clusters show the same properties as the visual star clusters.

\subsubsection{Mass Function of Visual Versus ML Star Clusters}\label{sec:massfn}
As the luminosity is an observed quantity, it can be quantified without any assumption of stellar models and/or SED fits. The luminosity function of young star clusters is generally described by a power law function $dN/dL \propto L^{-\alpha}$, with an almost universal index of $\alpha \sim 2$ as observed in local spiral galaxies \citep[e.g.,][]{larsen02, degrijs03}. Using the same method of \citet{messa18a}, we have computed the luminosity function of the visual star cluster catalogue and the ML star cluster catalogue between magnitudes of 18.5 to 23.25 on the V-band filter for the class 1 and 2 clusters of NGC~5194, finding that the clusters are well-described with a single power law fit with evidence for a steepening of the power law slope at the high-luminosity end, suggesting a truncation in the mass function \citep[see][]{haas08, messa18a}. The recovered single power law slopes are $1.96\pm0.04$ for the visual cluster catalogue and $1.92\pm0.03$ for the ML cluster catalogue. These are both consistent within the errors and demonstrates that the ML process does not penalize or mis-classify clusters with specific luminosities. The lower luminosity limit as reported in \citet{messa18a} of 23.25~mag is the same for both catalogues.

In order to explore the properties derived from SED fits to ensure we are not introducing biases in the ML catalogue, we calculate the mass function using the mass-complete sample (clusters with masses above 5000~M$_{\odot}$ and ages less than 200~Myr). The fit is performed both with a single power law and a truncated mass function; the latter serves to test the effect of random sampling from the mass function as such an effect can produce a truncation \citep[see][]{dasilva12}. For the single power law mass function, we recover slopes of $-2.12\pm 0.05$ for the visual cluster catalogue and $-2.18\pm0.02$ for the ML cluster catalogue. These slopes are consistent with the canonical $-2$ power law commonly seen in star cluster studies \citep[e.g.,][]{battinelli94, degrijs03, adamo17}.

\begin{figure*}
\includegraphics[scale=.42]{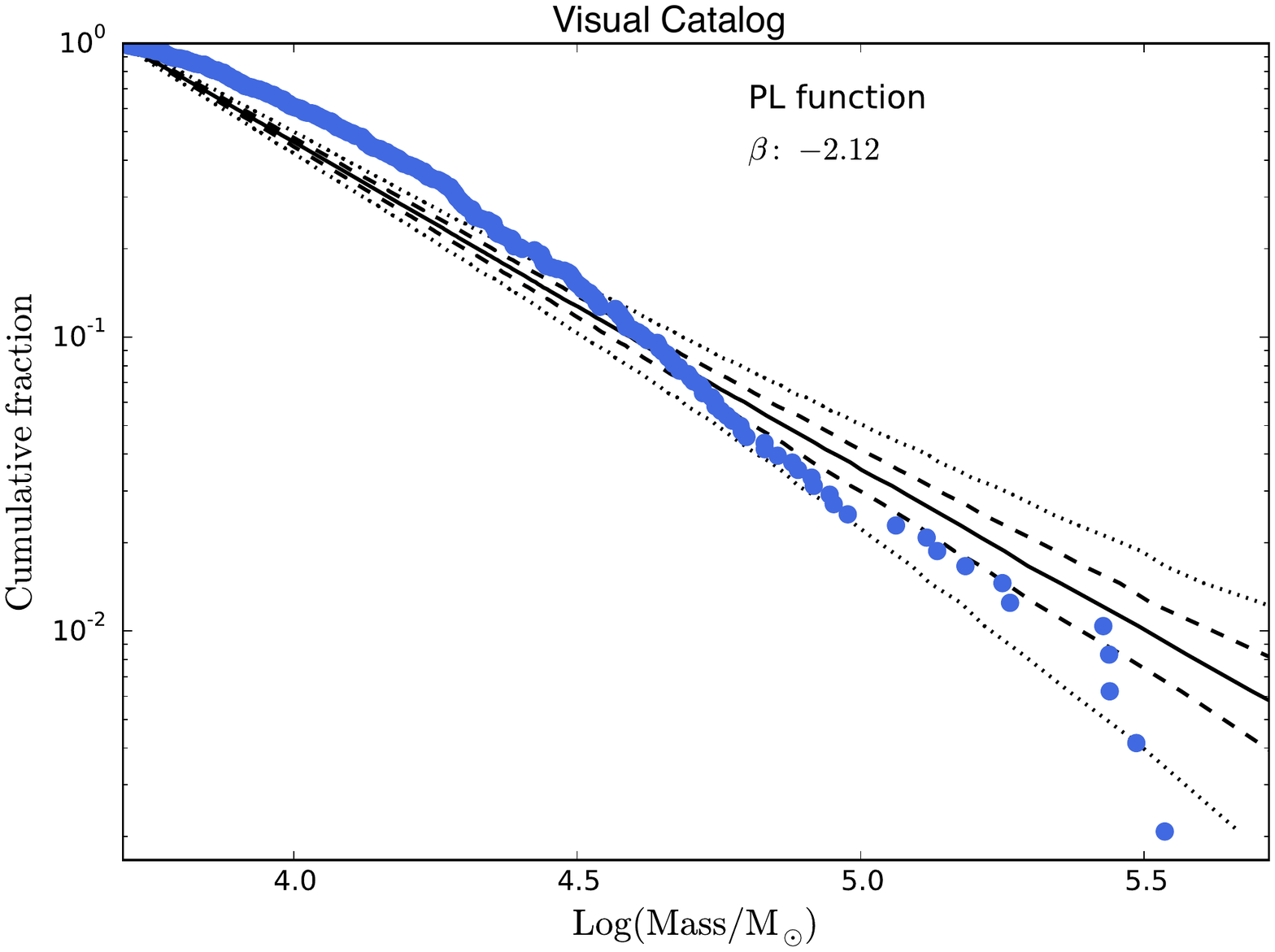}
\includegraphics[scale=.42]{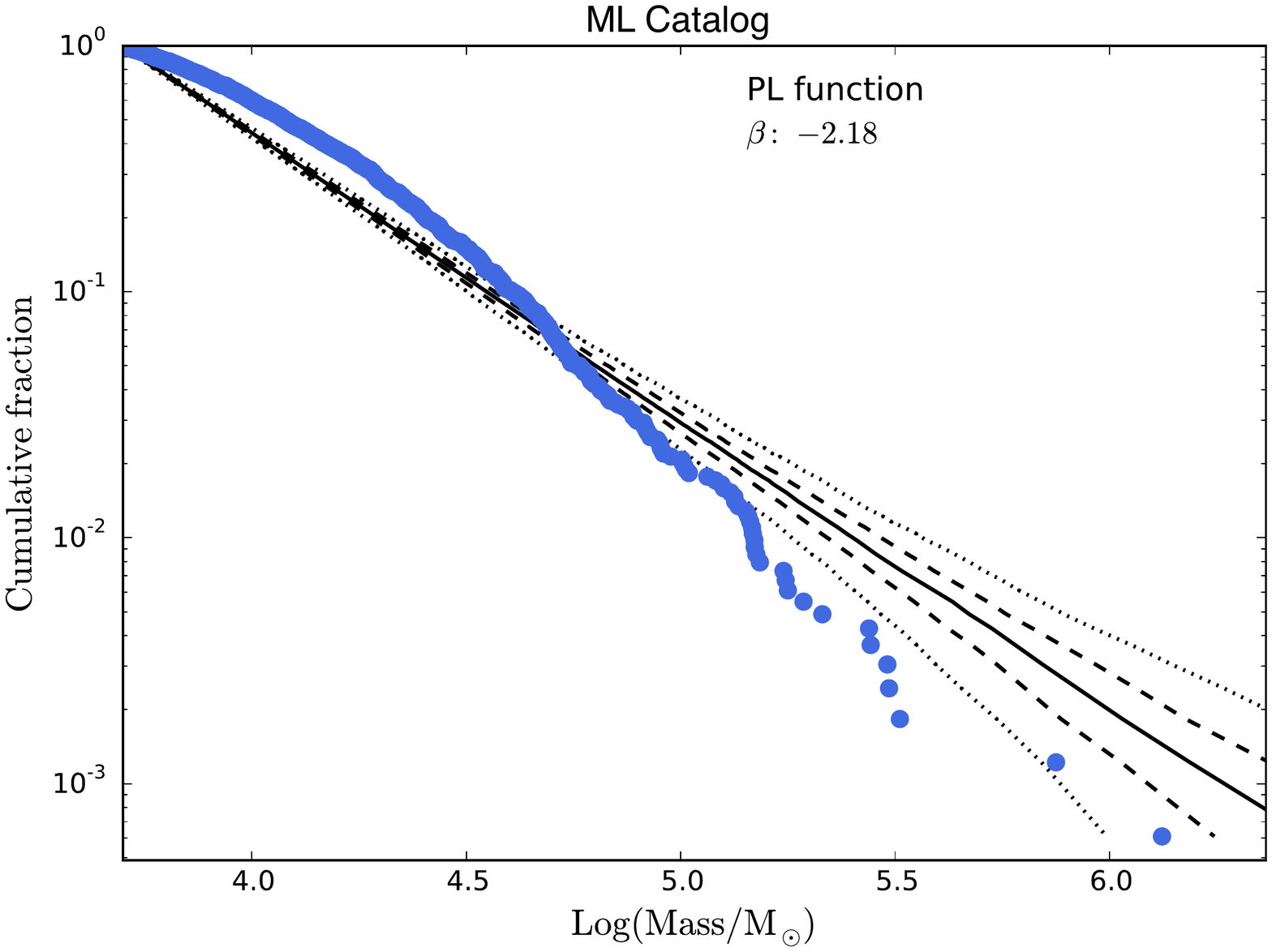} 
\includegraphics[scale=.42]{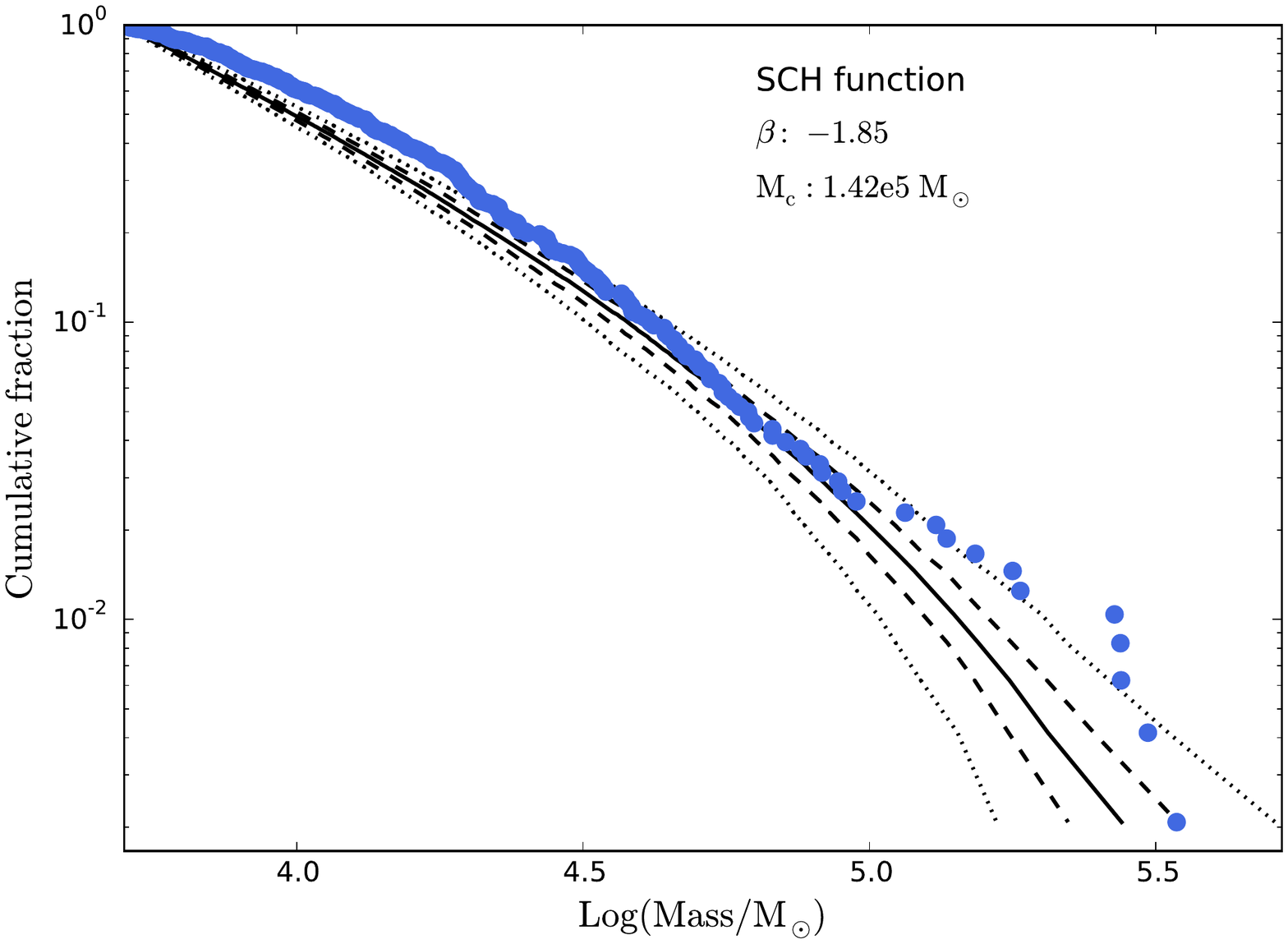} 
\includegraphics[scale=.42]{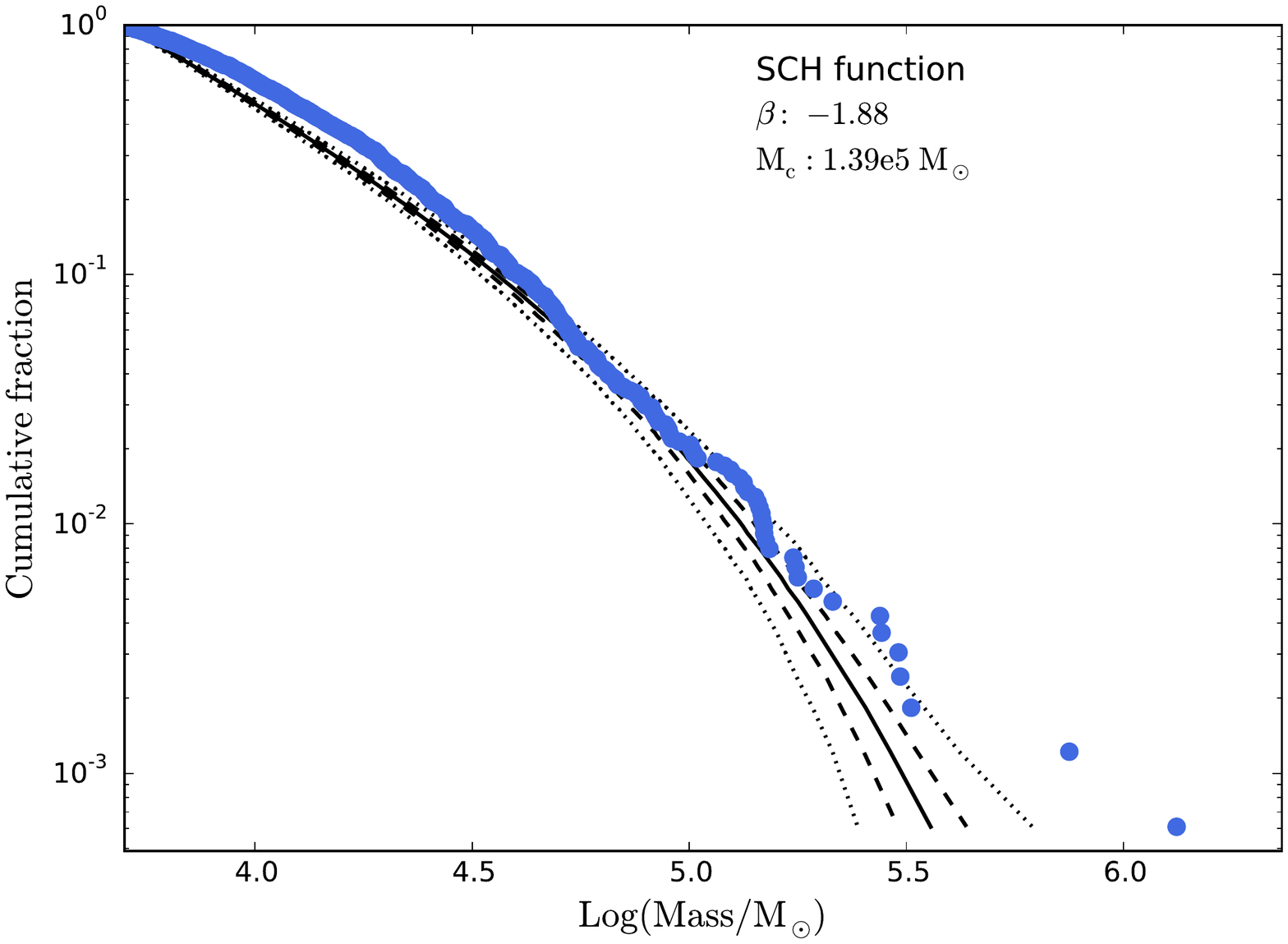} 
\caption{
Observed mass function (blue) for the visual catalogue (left column) and the ML catalogue (right column). Monte Carlo simulations are performed 1000 times and are shown as black lines. The median mass distributions (solid lines) and the limits within 50\% (dashed) and 90\% (dotted) of the simulations are plotted. The single power law fits and simulations are shown on the top row and the Schechter mass function are on the bottom. The differences observed between the two catalogues are consistent within the uncertainties.
\label{fig:massfunction}}
\end{figure*}

When fitted with a Schechter function with a truncation mass, we recover a slope of $-1.85 \pm 0.05$ with a truncation mass at $1.42 \pm 0.21 \times 10^5$~M$_{\odot}$ for the visual catalogue. The ML catalogue is best fit with a slope of $-1.88 \pm 0.03$ and a truncation mass of $1.39 \pm 0.14 \times 10^5$~M$_{\odot}$. We show the mass functions from both methods in Figure~\ref{fig:massfunction} along with simulated mass functions. The simulation mass functions are obtained via a bootstrapping technique from 1000 Monte Carlo trials as described in \citet{messa18a}. We show the median expectation, the 50\%, and the 90\% limits of the simulated function results. 

As the mass functions of ML and visual classified star cluster populations are consistent with each other within the uncertainties, we conclude the ML procedure is not introducing any bias in the mass parameter space of the clusters. 
%

\subsubsection{Age Function of Visual Versus ML Star Clusters}\label{sec:agefn}
\begin{figure}
\includegraphics[scale=.43]{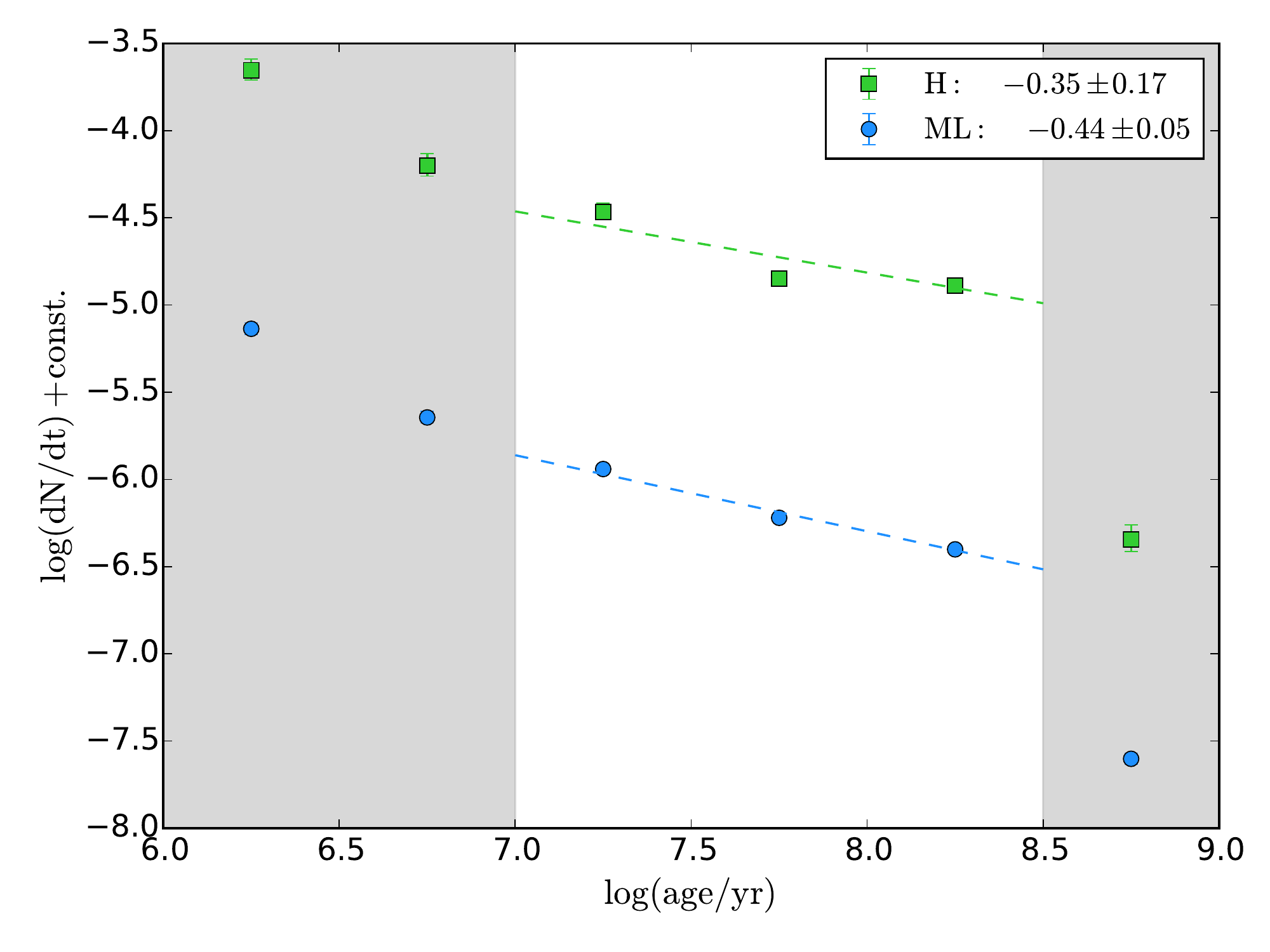} 
\vspace{-15pt}
\caption{
Age function of the visually identified star clusters (H: green) and the ML star cluster (blue). Dashed lines represent the best-fitting curve for the bins in the range log(age) = 7--8.5~yr. The grey-shaded areas mark the part of the functions excluded from the analysis due to incompleteness (old ages) and possible contamination by unbound sources (young ages). The differences in the slopes for the relations derived from the two catalogues are within the uncertainties. The visually identified data and relation have been shifted upward by 1.5~dex, for clarity. 
\label{fig:agefunction}}
\end{figure}

Our last check to ensure the robustness of our ML catalogue is to compare the age functions of the catalogues. We use the same method as implemented in \citet{messa18a} by dividing the sample in age bins of 0.5 and normalizing the number of sources in each bin by the age range spanned by each bin (Figure~\ref{fig:agefunction}). The points are fitted with a simple power law $dN/dt \propto t^{-\alpha}$ up to log(age/yr) = 8.5, where incompleteness starts to affect the results. We exclude sources in the fit of the age function that are younger than 10~Myr as they potentially contaminate our sample due to the rapid decline in the number of clusters surviving past $\sim$10~Myr \citep[see][]{lada03}. 

We recover a single power law fit to the age function for both classification method, with a slope of $-0.35 \pm0.17$ for the visual catalogue and $-0.44 \pm 0.05$ for the ML catalogue. For both methods, the differences are within the uncertainties. Our final conclusion is that the visual and ML catalogues are indistinguishable in terms of luminosity, age, and mass distribution properties of the identified star clusters.
%

\subsubsection{Contamination of Stars in the Cluster Catalogue}\label{sec:contamination}
\begin{figure*}
\includegraphics[scale=0.63]{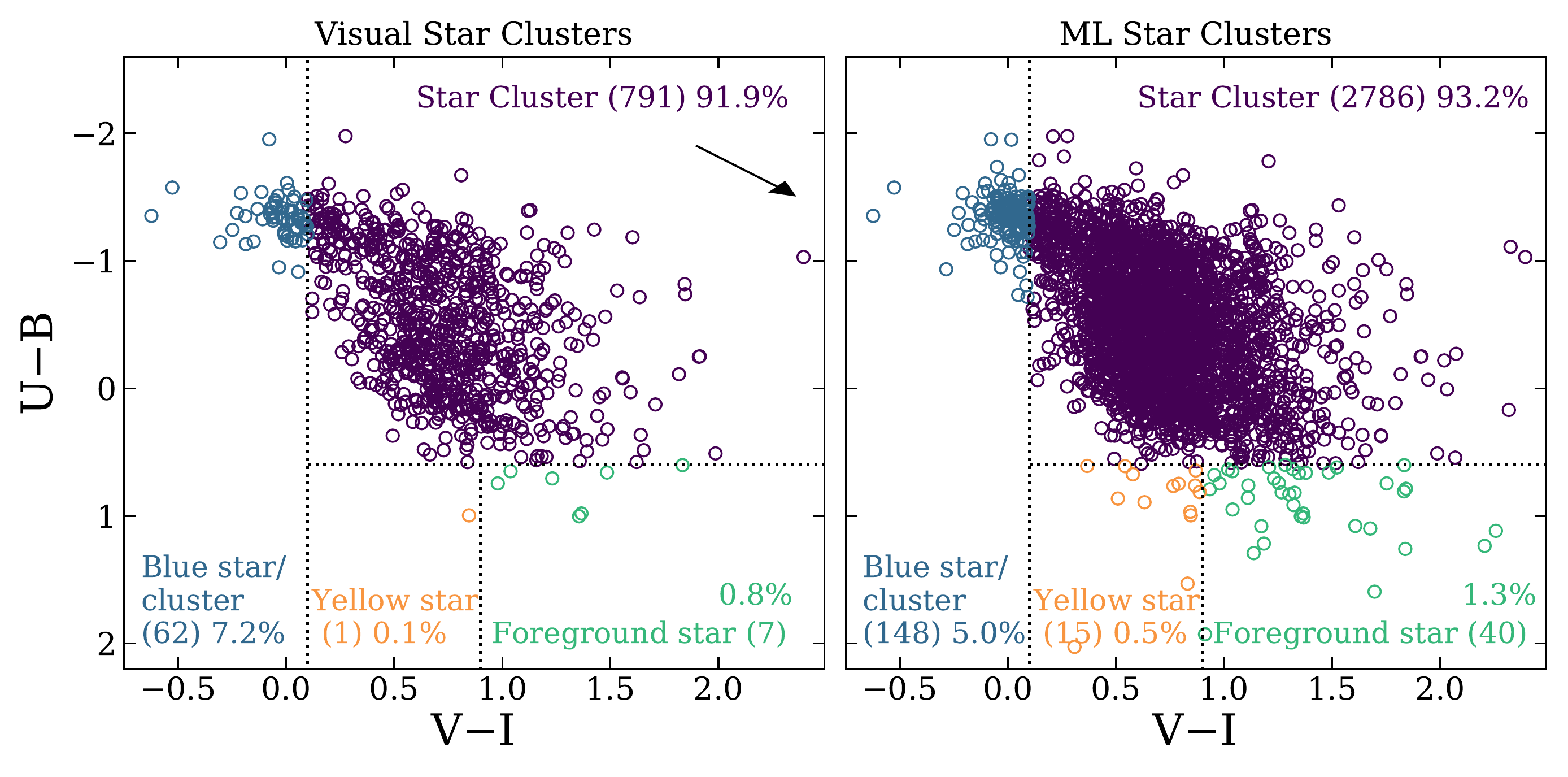}
\vspace{-15pt}
\caption{
U--B versus V--I colors for the star clusters with visual identification (left) and ML identification (right) for the star clusters in NGC~5194. The clusters are color coded by the different regions that deliminate in color space stars versus star clusters from \citet{whitmore10}. The numbers in parenthesis list the total number of sources within each region. We also list the fraction of each subregion compared to the total star cluster sample in each panel. The ML procedure does an exceptional job at identifying star clusters from individual stars and the contamination is only $\sim$2~\%, comparable to the contamination in the visually identified star cluster catalogue. The arrow marks reddening corresponding to $A_V$ = 1 mag (\ebv\ = 0.25 assuming a starburst attenuation curve).
\label{fig:colorcolor}}
\end{figure*}

Figure~\ref{fig:colorcolor} shows the distribution in color-color space that separates single stars from star clusters, using the method by \citet{whitmore10}. In our catalogues, the color-color space occupied by single stars and clusters is well-separated and the contamination rate of individual stars in the ML cluster catalogue is low, $\sim$2\%. This demonstrates that our visual classification procedure, as well as the ML procedure, is successful in differentiating between star clusters and contaminants. However, color-color cuts alone do not suffice to differentiate star clusters from stars and the class 4 non-cluster contaminants occupy much of the same space as the star clusters \citep{whitmore10}. The inability to discriminate between faint clusters and massive stars necessitate the addition of cuts in the size (concentration index) as well as visual morphology to create robust and reliable star cluster catalogues. 

%

\section{Results and Analysis}\label{sec:results}

\subsection{Correlating the Young Star Clusters to Molecular Gas}\label{sec:gas}
\begin{figure}
\includegraphics[scale=0.42]{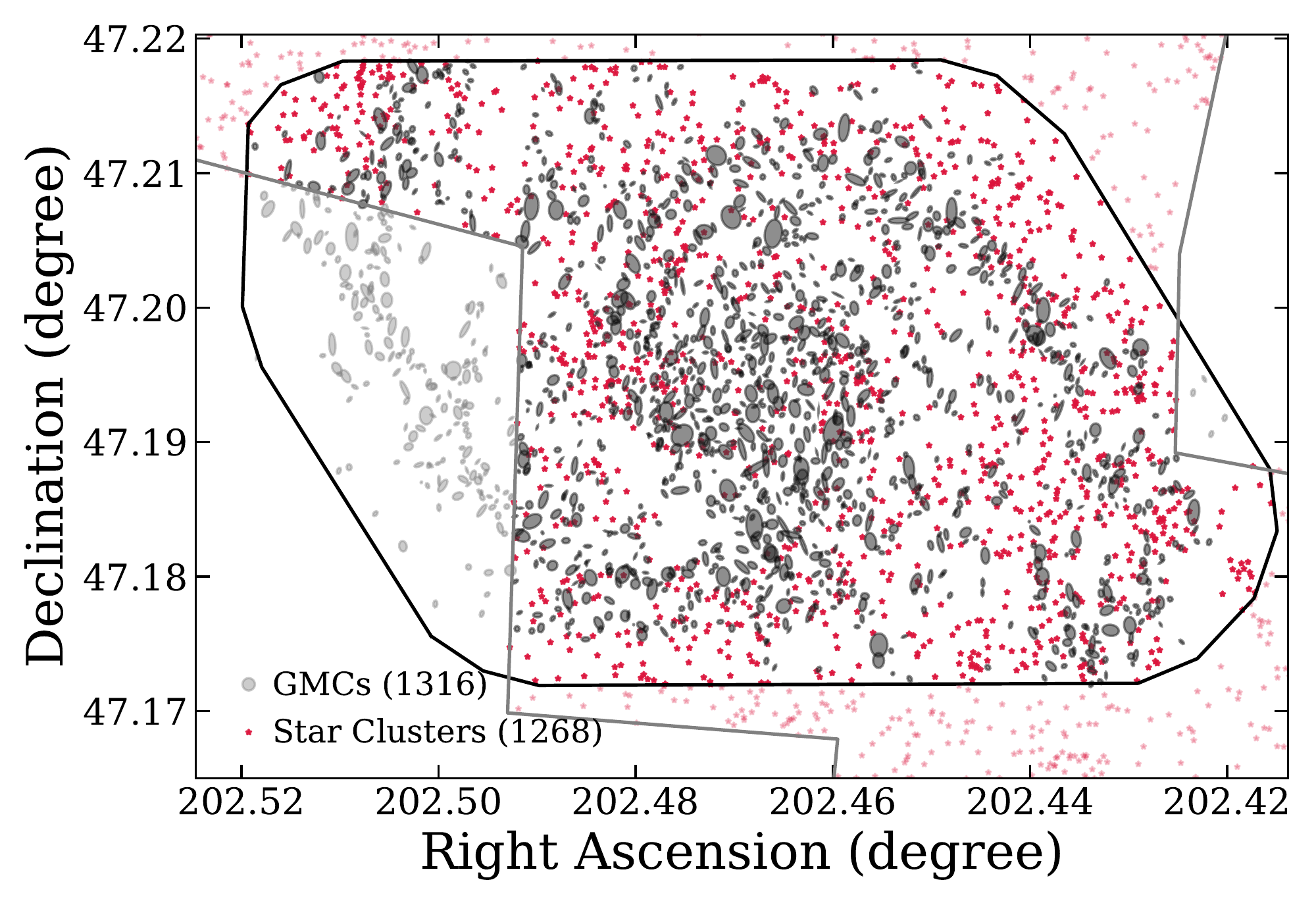}
\caption{
Location and sizes of the GMCs and star clusters. The black line shows the outline of the PAWS coverage for the GMCs and the gray line shows the UVIS coverage for the star clusters. The numbers listed are the star clusters and GMCs within each footprint. Star clusters and GMCs that are not located within the PAWS and UVIS footprint are shown in light pink and light gray, respectively, and excluded from all star cluster--GMC comparisons. 
\label{fig:SC_GMC}}
\end{figure}

To connect the young star clusters to their environment, we compare the cluster locations to the location of molecular gas. We use the GMC catalogue \citep{colombo14a} from the Plateau de Bure Interferometer Arcsecond Whirlpool Survey \citep[PAWS;][]{schinnerer13,pety13}. PAWS mapped the molecular gas in the central 9 kpc of NGC~5194 using the $^{12}$CO(1--0) line emission at a cloud-scale resolution of $\sim$40~pc. The GMCs identified by Colombo et al. (2014) account for about half (54\%) of the flux in the PAWS map. The remainder lies in structures that are blended, continuous, diffuse, low S/N or otherwise intractable by their cloud-finding approach. By construction, the GMCs represent significant peaks in the CO emission, and assuming that these correspond to the cluster-forming structures seems reasonable. Figure~\ref{fig:SC_GMC} shows the inner region with the star clusters and GMCs that lie within the footprint coverage of both UVIS and PAWS, for a total of 1268 class 1 and 2 star clusters and 1316 GMCs. 
%

Previous observations show that stars and star clusters quickly become unassociated with the GMCs from where they are born, either due to drift or from blowing out the gas. The timescale for most clusters to become exposed and lose association with their molecular gas is as early as 2~Myr within M83 \citep{hollyhead15} to $\sim$5~Myr from studies of the Antennae galaxy \citep{whitmore14}. We perform multiple tests to correlate the location of star clusters with the molecular clouds to investigate if star clusters become less associated with molecular clouds with age.

\subsubsection{Distance of Star Clusters from Molecular Clouds}
\begin{figure}
\includegraphics[scale=0.48]{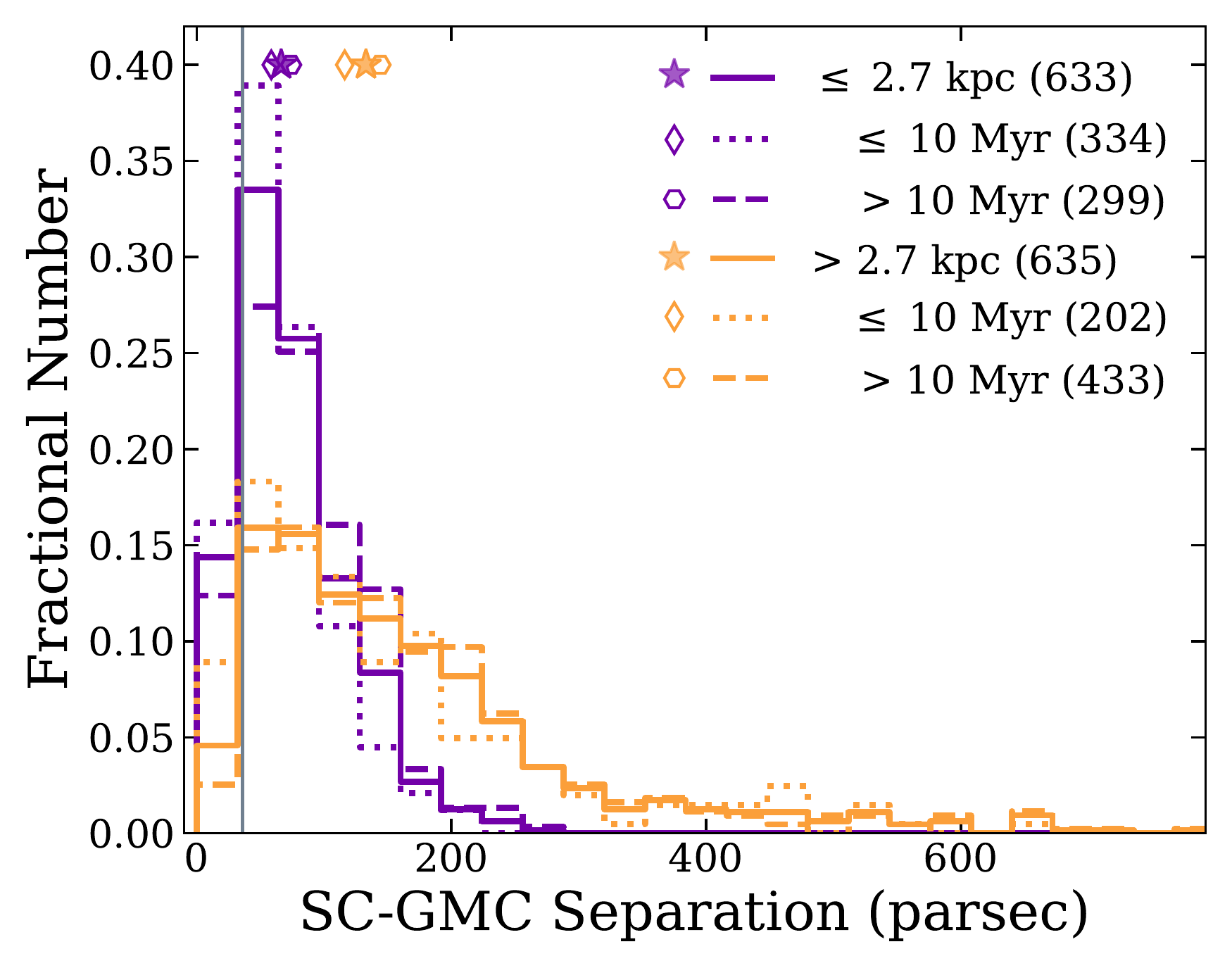} 
\caption{
Fractional distribution of the shortest distance for each star cluster to the closest GMC broken into two radial bins and age bins. The symbols show the median value for each distribution. Star clusters less than a galactocentric radius of 2.7~kpc (purple) are $66\pm2$~pc from the nearest GMC, with ages less than 10~Myr (dotted line) showing slightly shorter distances of $59\pm2$~pc compared to ages greater than 10~Myr (dashed line) at $74\pm4$~pc. Star clusters at galactocentric distances greater than 2.7~kpc (orange) are $132\pm6$~pc from the nearest GMC, with distance dropping for the youngest ($\leq$10~Myr; dotted line) to $118\pm9$~pc and increasing to $143\pm7$~pc for the older clusters ($>$10~Myr; dashed line). On average, younger star clusters are closer in proximity to a GMC than older star clusters although the distance from the center of the galaxy has a bigger impact, with the average SC--GMC distance increasing with increasing galactocentric distance. The 1$\sigma$ errors are bootstrap estimates based on 10,000 samples. The vertical line shows the median GMC radius of 36~pc.
\label{fig:sc_gmc_mindistance}}
\end{figure}

In our first test, we take the star clusters within the PAWS coverage and calculate the distance to the center of their nearest GMC (Figure~\ref{fig:sc_gmc_mindistance}). For the entire sample, the median of the shortest distance for the closest star cluster--GMC pair is $87\pm2$ parsec. For comparison, the median diameter of GMCs in this galaxy is 72~pc. The 1$\sigma$ errors are bootstrap estimates based on 10,000 samples. We divide the star cluster sample into half, corresponding to a galactocentric distance of 2.7 kpc, to investigate if distance from the galaxy center or age of the clusters is the primary driver in the change of distance between pairs of star clusters and GMCs. Star clusters at galactocentric distance less than 2.7~kpc are substantially closer to their nearest GMC at $66\pm2$~pc whereas the shortest distance between all cluster--GMC pair doubles to $132\pm6$~pc at galactocentric distances greater than 2.7~kpc. We further separate the two radial bins into two age bins, older and younger than 10~Myr. We find that for galactocentric distances less than 2.7~kpc, the younger clusters are separated from their nearest GMCs by $59\pm2$~pc, and the older clusters are separated from their GMCs by $74\pm4$~pc. Similarly, at galactocentric distances larger than 2.7~kpc, the younger and older clusters are separated from their nearest GMCs by $118\pm9$~pc and $143\pm7$~pc, respectively.

When only considering age, the star clusters younger and older than 10~Myr exhibit median distances of $71\pm3$~pc and $107\pm4$~pc, respectively, which are different separation at a 7--$\sigma$ level significance. The median radius of the GMC population in NGC~5194 is 36~parsec, thus star clusters that are younger than 10~Myr lie at distances twice the size of the cloud's radius whereas clusters older than 10~Myr are found at median distances three times the size of the molecular cloud. 

While on average younger star clusters are closer in proximity to their nearest GMC, the galactocentric distance is a stronger predictor of the trend for increasing distance between clusters and GMCs. Both the aggregate midplane pressure \citep{elmegreen89} and the thermal gas pressure \citep{wolfire03} are expected to fall with radius in disk galaxies. Both the surface density and velocity dispersion of the gas are observed to decrease with radius in NGC~5194 \citep[e.g.,][]{schuster07, tamburro09}, as does the mean volume density of the gas \citep{chen15, bigiel16}. In this case, one explanation for our observation is that in the lower density and environment at larger galactocentric radii, the number density of clusters and molecular clouds is lower than at small galactocentric distances, leading to a lower chance of young clusters to be associated with their parent cloud. Star clusters and their unbound complexes do respond to the environment that they live in; we find that clusters in the outer region of the galaxy reside in complexes significantly larger than clusters closer to the galaxy center, an effect of the increased shear, midplane pressure, and turbulence. GMCs are also more effectively dispersed within the inner region of NGC~5194 due to the increased shear \citep{meidt15}. We further investigate the effect of the environment on star-forming complexes in Section \ref{sec:radial}. These results do not inform us if a star cluster is actually associated with the closest GMC; we investigate this more in the next section.
%

\subsubsection{Star Clusters Associated with GMCs}\label{sec:SCs_insideGMCs}
\begin{figure}
\includegraphics[scale=0.42]{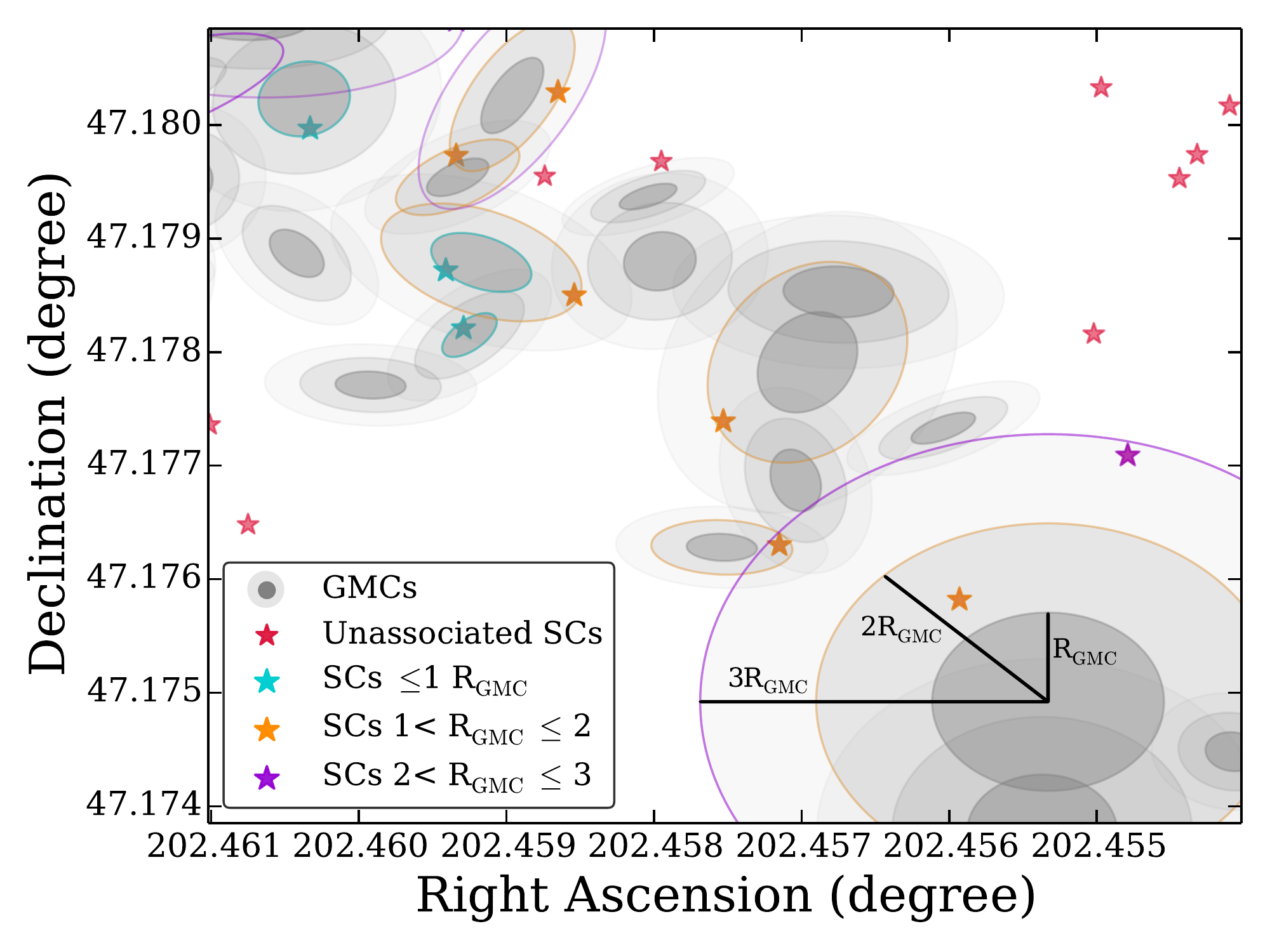}
\caption{
Associated star clusters located inside 1 $R_{\rm GMC}$ (turquoise), 1--2 $R_{\rm GMC}$ (orange), and 2--3 $R_{\rm GMC}$ (purple) while unassociated star clusters are shown in pink. We consider star clusters to be associated with a GMC if they lie within the footprint of the molecular cloud. GMCs that have a corresponding cluster are outlined in the same color as the radial location of the cluster. 
\label{fig:scinsidegmc}}
\end{figure}

In our next test, we examine the properties of the young star clusters that are still associated with a molecular cloud. We define the association of a star cluster and a GMC as cases where the position of the star cluster falls within the footprint of the GMC. Likewise, a cluster becomes unassociated (i.e., ``detached'') from its parental GMC when it has drifted away from its natal environment. In addition to detaching via drift, clusters are capable of photo-evaporating and blowing away via stellar feedback/winds the ambient molecular gas; these cases reveal themselves as holes in the GMC but are indeed still associated with the ambient molecular gas \citep[see, e.g.,][]{hony15}. We treat the GMCs as ellipses and even if there are clusters that have evacuated their surrounding material, blowing out holes in the adjacent gas, these cases would still be considered associated, since they fall within the footprint of the GMC ellipse. Figure~\ref{fig:scinsidegmc} shows an example of star clusters that are located within the $1 R_{\rm GMC}$ ellipse footprint of a GMC in a zoom in of the galaxy. We repeat the analysis for star clusters within an annulus between 1 and $2 R_{\rm GMC}$, between 2 and $3 R_{\rm GMC}$, and those beyond $3 R_{\rm GMC}$ distance. 

We expect that if star clusters do disperse from their GMCs with time, clusters that lie further than $3 R_{\rm GMC}$ will be older than those within $1 R_{\rm GMC}$. When counting the star clusters that lie within a GMC, we only allow each cluster to be assigned to a single GMC. In situations where multiple clouds lie on top of each other and a star cluster can potentially be paired up with more than one cloud, the star cluster is assigned to the most massive GMC. If we instead assign a star cluster to the closest GMC regardless of its mass, the results remain robust as only 3\% percent of the star clusters lie on top of more than one cloud, and in many situations, the closest cloud is still often the most massive as well (see further discussion in Section \ref{sec:massivegmcs}). We perform the search in order of increasing distance from any GMC such that all star clusters that reside within $1 R_{\rm GMC}$ are assigned to their associated cloud before moving onto star clusters that reside in the annuli between 1 and 2 $R_{\rm GMC}$ and 2 and 3 $R_{\rm GMC}$, respectively. 

\begin{figure*}
\includegraphics[scale=0.43]{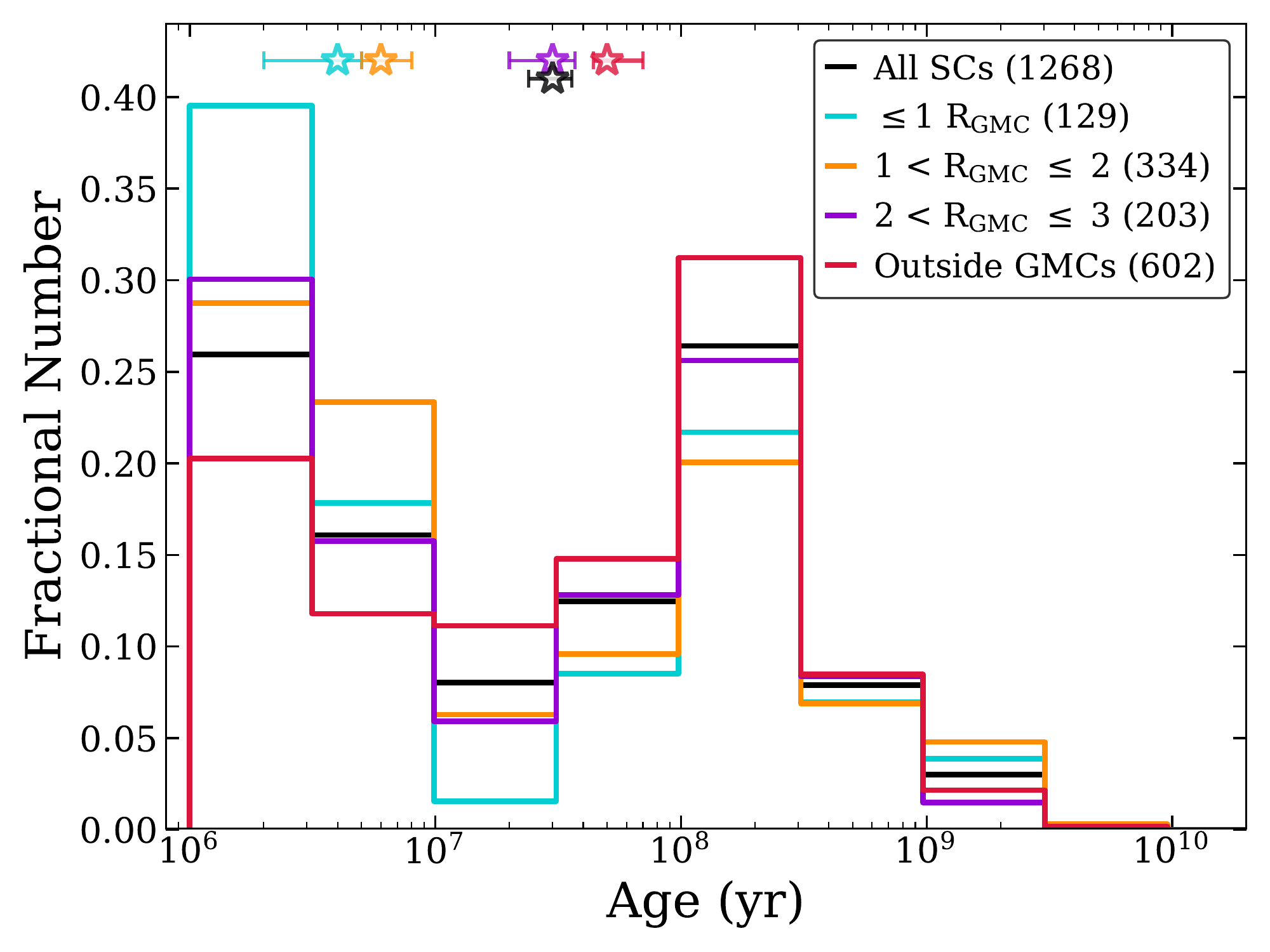}
\includegraphics[scale=0.43]{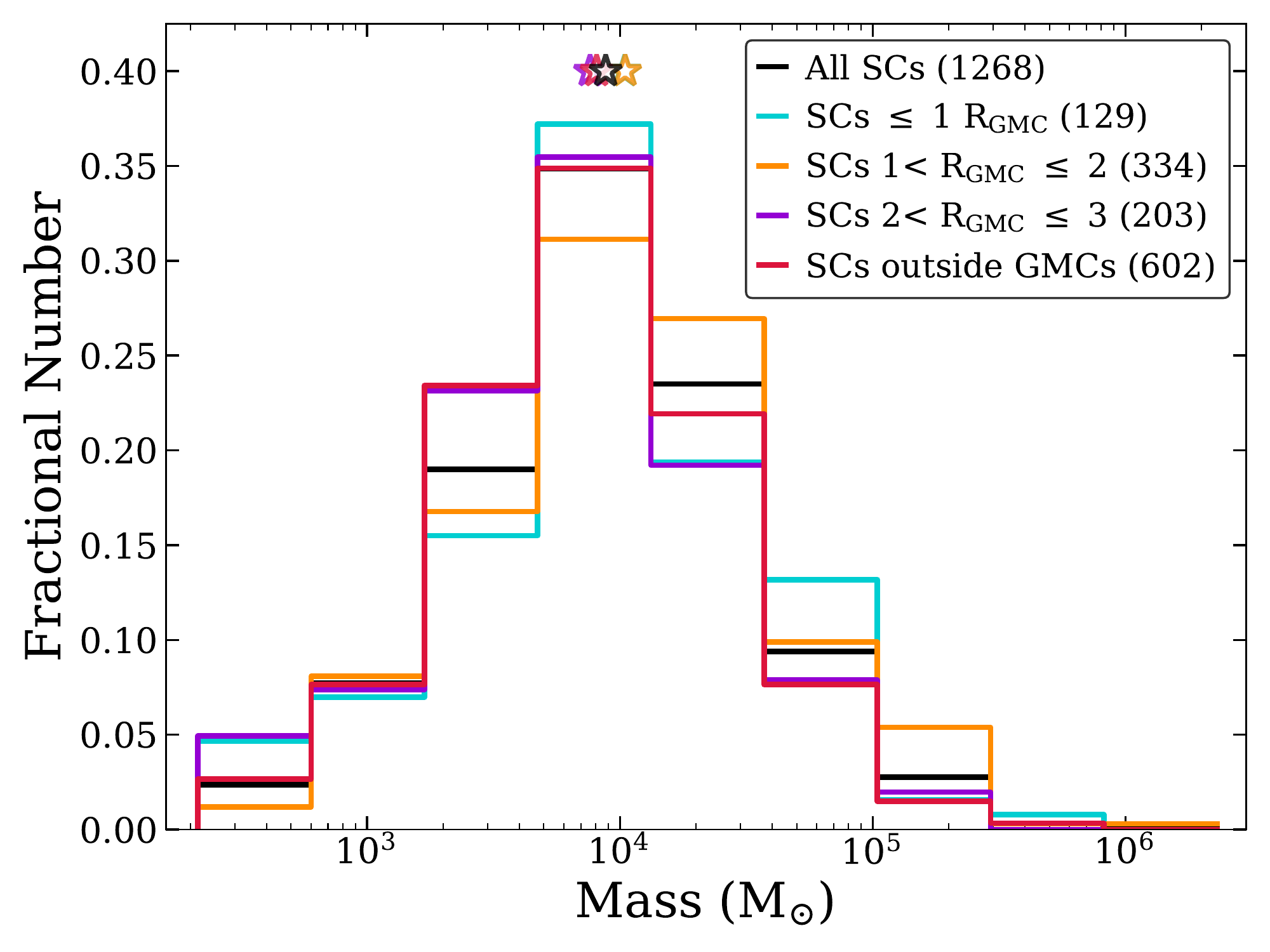}
\caption{
{\bf Left:} Normalized distribution of the ages of star clusters for the entire sample (black), star clusters located $\leq$1 $R_{\rm GMC}$ (turquoise), star clusters within 1--2 $R_{\rm GMC}$ (orange), 2--3 $R_{\rm GMC}$ (purple), and star clusters unassociated with a GMC ($>$3 $R_{\rm GMC}$; red). The stars show the median age of each distribution (30$^{+6}_{-6}$~Myr, 4$^{+1}_{-2}$~Myr, 6$^{+2}_{-1}$~Myr, 30$^{+7}_{-10}$~Myr, and 50$^{+20}_{-10}$~Myr, respectively) and the 1$\sigma$ errors on the median are bootstrap estimates based on 10,000 samples. Star clusters located within a GMC are generally much younger than star clusters that are further away from a GMC and the age progressively increases with increasing distance from the GMC. 
{\bf Right:} Normalized masss distribution of the star clusters. The stars show the median mass of each distribution ($8780\pm840~M_{\odot}$, $10450\pm1090~M_{\odot}$, $10490\pm940~M_{\odot}$, $7620\pm820~M_{\odot}$, and $8090\pm910~M_{\odot}$). The mass distribution of star clusters located within a GMC (turquoise) is statistically the same as the mass distribution of star clusters that are unassociated with a GMC (red). The 1$\sigma$ errors from bootstrap estimates based on 10,000 samples are not shown in the plot as the uncertainties are $\sim$10\% of the median value. 
\label{fig:dist_agemass}}
\end{figure*}

We find a total of 129 star clusters that lie within the footprint of 112 GMCs ($\leq 1 R_{\rm GMC}$). An additional 334 clusters are between 1 and 2 radii of a GMC center and 203 clusters are between 2 and 3 radii of a GMC, leaving 602 star clusters unassociated with any cloud ($\geq 3 R_{\rm GMC}$). Figure~\ref{fig:dist_agemass} shows the distribution of the cluster age and masses. The median age of all clusters in NGC~5194 is 30$^{+6}_{-6}$~Myr, significantly older than clusters located within one radius of a GMC with median ages of 4$^{+1}_{-2}$~Myr. Of the clusters still within a GMC radius, nearly 40\% have ages less than 3~Myr. The median age rises slightly for clusters between one and two radii of a GMC with a median age of 6$^{+2}_{-1}$~Myr, and clusters that are between 2 and 3 radii from a GMC have median ages of 30$^{+7}_{-10}$~Myr. Star clusters that are unassociated with any GMC have median ages of 50$^{+20}_{-10}$~Myr. The 1$\sigma$ uncertainties are bootstrap estimates based on 10,000 samples; the age trend is significant at the 6--$\sigma$ level. 

There is an insignificant difference in the cluster mass with respect to its distance to the nearest GMC: the median cluster mass is $10450\pm1090$, $10490\pm940$, $7620\pm820$, and $8090\pm910~M_{\odot}$ for clusters within 1 $R_{\rm GMC}$, 2 $R_{\rm GMC}$, 3 $R_{\rm GMC}$, and unassociated, respectively. Table~\ref{tab1} lists the properties of the star clusters and, in addition, is divided by the region --- center, spiral arm, and interarm --- of the galaxy as defined by the PAWS dataset \citep{colombo14a}. We are limited by incompleteness within the very center of the galaxy due to the high background and we do not detect sources older than $\sim$10~Myr due to the blending of older sources with similar color to the background light. The average completeness in the center of the galaxy is 1.08~mag higher than in the galactic disk, a difference of 0.42~mag in the UV-band, 0.96~mag in the U-band, and 1.52~mag in the B-band, 1.39~mag in the V-band, and 1.09~mag difference in the R-band \citep{messa18a}. The cluster catalogues of \citet{chandar17} also show a dearth of clusters in the inner $\sim$1~kpc of the galactic center. This appears to not affect the clusters in the central GMC region that are within 1 $R_{\rm GMC}$ where the completeness age is typically older than the age of clusters that are still associated with a GMC ($\sim$4~Myr), though it begins to impact the results for clusters with distances greater than 2 $R_{\rm GMC}$ as those systems are typically older than the completeness age within the center. 

We find average \ebv\ values of 0.21, 0.22, 0.24, and 0.25 for clusters within 1 $R_{\rm GMC}$, 2$R_{\rm GMC}$, 3$R_{\rm GMC}$, and unassociated, respectively. For a starburst attenuation curve, these correspond to $A_V$'s of 0.85, 0.89, 0.96, and 1.0, respectively. The scatter in \ebv\ for a given age range is significant and it does not correlate strongly with age \citep{messa18a}. Clusters closer to a GMC do not appear to be any more embedded than those further from a GMC, as indicated by their \ebv\ values. The difference in age between star clusters inside/outside GMCs therefore cannot be explained by significantly higher extinction affecting the star clusters within the spiral arms or GMCs. If anything, we would expect the youngest clusters to be embedded within the clouds, and therefore not to be recovered by our inspections that rely on detection of UV/blue emission from the clusters. This consideration further reinforces the significance of our result, that the youngest clusters are recovered in proximity of GMCs.

The trend for younger clusters to lie near GMCs is anticipated as 4~Myr old stellar systems are expected to have already evacuated their surrounding cloud material \citep{hollyhead15}, but have not lived long enough to have traveled far enough to become separated with their birth location. The ages we recover here are younger than typical GMC dissolution timescales of $\sim$10--30~Myr from both observations and theoretical expectations \citep[e.g.,][]{engargiola03, murray11, dobbs13, heyer15, meidt15, jeffreson18}. We can calculate the traveling velocity of the star clusters at distances less than 3~$R_{\rm GMC}$, assuming they originate from the center of the nearest cloud, by dividing the shortest distance between each star cluster -- GMC pair by the measured age of the star cluster. We find that the average velocity required for a star cluster to travel to its current location relative to the center of the nearest GMC is 9.5~\kms. This velocity is a result of ionization which erodes the cloud near the cluster also combined with actual motion of the near-edge of the GMC in an expanding HII region, and motions induced by dynamical effects in addition to drift of the cluster as well. These results provide direct measurements of the age and velocity necessary for the dissociation of clusters from their natal clouds, in agreement with the ISM velocities required to randomize star clusters from their hierarchical complexes (see Section \ref{sec:2pcf} and \ref{sec:discussion}).  
%
%
\begin{table}
 \caption{Properties of star clusters depending on their association with a GMC or location within different regions of the galaxy as defined in the PAWS dataset \citep{colombo14a}. Columns list the:
(1) location of the star cluster; 
(2) number of GMCs; 
(3) number of star clusters; 
(4) median age of the star clusters; and
(5) median mass of the star clusters. The 1$\sigma$ uncertainties are bootstrap estimates based on 10,000 samples. Numbers in parentheses indicate uncertainties in the final digit(s) of listed quantities, when available. }
\label{tab1}
 \begin{tabular}{lcccc}
  \hline
Region 	& 
N$_{\rm GMC}$  	& 
N$_{\rm SC}$ 	& 
Median Age & 
Median Mass \\
& & & (Myr) & (M$_{\odot}$) \\
  \hline
Within 1$R_{\rm GMC}$		& 112	& 129	& 4$^{+1}_{-2}$ 		& 10450(1090) \\
\hspace{10pt}Center 		& 33 	& 36		& 4$^{+1}_{-1}$ 		& 10830(1510) \\
\hspace{10pt}Spiral Arm	& 51		& 57		& 4$^{+2}_{-2}$ 		& 11190(2480) \\
\hspace{10pt}Inter Arm		& 28		& 36 	& 3$^{+2}_{-1}$ 		& 7559(2280) \\ \\
Within 2$R_{\rm GMC}$	& 242	& 334	& 6$^{+2}_{-1}$ 		& 10490(940) \\
\hspace{10pt}Center 		& 57 		& 86		& 4$^{+0.5}_{-1}$		& 13680(2020) \\
\hspace{10pt}Spiral Arm	& 96		& 135	& 15$^{+8}_{-5}$		& 10920(1560) \\
\hspace{10pt}Inter Arm		& 89		& 113 	& 7$^{+2}_{-4}$		& 8100(1140) \\ \\
Within 3$R_{\rm GMC}$	& 168	& 203	& 30$^{+7}_{-10}$	& 7620(820) \\
\hspace{10pt}Center 		& 31		& 37		& 3$^{+0.5}_{-0.5}$	& 7280(1310) \\
\hspace{10pt}Spiral Arm	& 75		& 79		& 50$^{+12}_{-9}$	& 8710(1610) \\
\hspace{10pt}Inter Arm		& 62		& 87		& 50$^{+10}_{-6}$	& 7790(960) \\ \\
Unassociated 					& 934	& 602	& 50$^{+20}_{-10}$	& 8090(910) \\
\hspace{10pt}Center 		& 245	& 25		& 4$^{+3}_{-2}$ 		& 8850(2760) \\
\hspace{10pt}Spiral Arm	& 385	& 122	& 50$^{+13}_{-15}$	& 11680(1440) \\
\hspace{10pt}Inter Arm		& 304	& 455	& 50$^{+6}_{-8}$		& 7300(430) \\ \\
Total 								& 1316	& 1268	& 30$^{+6}_{-6}$		& 8780(840) \\ 
\hspace{10pt}Center 		& 335	& 184	& 4$^{+1}_{-0.5}$		& 10660(1140) \\
\hspace{10pt}Spiral Arm	& 540	& 393	& 50$^{+8}_{-7}$		& 11130(870) \\
\hspace{10pt}Inter Arm		& 441	& 691 	& 50	$^{+5}_{-5}$		& 7540(720) \\
  \hline
 \end{tabular}
\end{table}

\subsubsection{Are Massive GMCs more likely to Host Multiple Star Clusters?}\label{sec:massivegmcs}
\begin{figure*}
\includegraphics[scale=0.47]{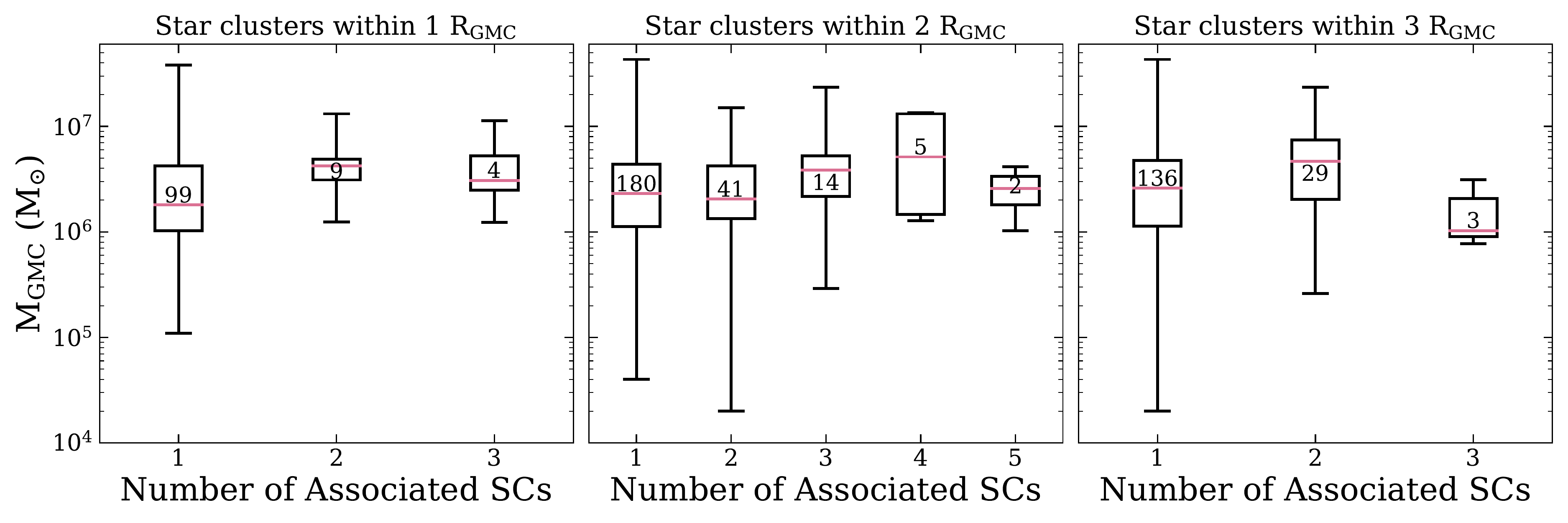}
\caption{
Boxplots of the GMC masses versus the number of associated star clusters. The numbers listed in the middle of each box show the total number of each GMC for a given number of associated star clusters and the pink line shows the median of the GMC masses. The box encloses 50\% of the distribution (the first and third quartile) and the whiskers extend to the minimum and maximum. The left figure shows the GMC mass distribution for the 118 GMCs with 129 star clusters within one radius of the cloud, the middle panel shows the mass distribution for the 242 GMCs with 334 star clusters between and and two radii of each cloud, and the right panel shows the mass distribution for the 168 clouds with 203 star clusters between two and three radii. 
\label{fig:massGMCvsnumSC}}
\end{figure*}

We also examine if more massive GMCs are statistically more likely to host multiple star clusters. Figure~\ref{fig:massGMCvsnumSC} shows the distribution of the mass of GMCs as a function of number of star clusters located within its footprint, for both star clusters found within 1 $R_{\rm GMC}$ and those found within 2 $R_{\rm GMC}$. There is a marginal statistically insignificant trend for more massive GMCs to host multiple star clusters. As the number of associated star clusters increases, the total spread in the mass of the host GMC does decrease. 

Star clusters that lie on top of multiple GMCs can potentially impact this result as we always assign a cluster to the most massive GMC in such an event. However, there are only three star clusters (3\%) within one radius of a GMC that have the possibility of being paired up to more than one cloud. For all three cases, these star clusters were the only clusters associated with either GMC. Changing the associated GMC for these clusters away from the most massive GMC only moves that data points downward and both the median and the quartiles remain unchanged (Figure~\ref{fig:massGMCvsnumSC}a). For the star clusters within two radii of the center of a cloud, 48 (14\%) lie ontop of two GMCs and three clusters (0.9\%) lie on top of three clouds. We randomly assign these clusters to different clouds and recompute the values in Figure~\ref{fig:massGMCvsnumSC}b and we find that the effect is minimal and that the median value always falls within the first and third quartiles. Additionally, we find no evidence for more massive GMCs to be hosts of more massive star clusters. 
%

\subsection{The Two-Point Correlation Function}\label{sec:2pcf}
We implement the angular two-point correlation function $\omega(\theta)$ to measure the magnitude of clustering as a function of projected distance between the star clusters. A detailed description of the formalism and methodology of the two-point correlation function as applied to star clusters within other LEGUS galaxies can be found in both \citet{grasha15} and \citet{grasha17a}. The correlation function provides a way to identify common age structures to derive the randomization timescale for when the coherence of star-forming hierarchies becomes uncorrelated (Section \ref{sec:agecuts}) as well as the correlation length --- the size scale for the star-forming regions --- as a function of age and location within a galaxy (Section \ref{sec:radial}). Here we list the details necessary for the application to the star clusters and the GMCs within NGC~5194. 

The correlation function \citep{peebles80} is defined as a measure of the probability of finding a neighboring object, above what is expected for an unclustered random Poisson distribution. In this study, we implement the angular two-dimensional correlation function $\omega(\theta)$, as projected onto the plane of the sky, as the probability above Poisson of finding two star clusters within an angular separation $\theta$ as $\mathrm{d}P = N^2 [ 1 +  \omega(\theta)]\ \mathrm{d}\Omega_1 \mathrm{d}\Omega_2$, where $N$ is the surface density of clusters per steradian with two infinitesimal elements of solid angle $\mathrm{d}\Omega_1$ and $\mathrm{d}\Omega_2$, separated by angle $\theta$. 

To calculate $\omega(\theta)$, we count pairs of star clusters as a function of separation and compare that to what is expected for an unclustered distribution. A clustered distribution has an excess of pair counts at small separations, resulting in $\omega(\theta)>0$ at small length scales whereas random distribution of an unclustered population results in a flat correlation with $\omega(\theta) = 0$ at all length scales. We supplement the cluster data with a catalogue of random sources that populates the same sky coverage and geometry (e.g., edges, masks) as the real data. The ratio of pairs of clusters observed in the data relative to pairs of points in the random catalogue is then used to estimate $\omega(\theta)$ with the \citet{landy93} estimator. We fit the projected angular two-point correlation function with a power-law,
\begin{equation}\label{eq:powerlaw}
1+\omega(\theta) = A_{\omega}\theta^{\alpha},
\end{equation}
where the slope $\alpha$ measures the strength of the clustering and the amplitude $A_{\omega}$ measures the correlation length of the clustering; we use both to determine if the clustering is consistent with being scale-free. We determine the exponent $\alpha$ of the correlation function through a linear regression on the log--log plots of all of our correlation functions by applying a Levenberg--Marquardt non-linear least-squares minimization fit. 

The original work of \citet{peebles80}, along with the application to stellar populations by \citet{gomez93} and \citet{zhang01}, fits the power law distribution to $\omega(\theta)$. Following the convention of \citet{scheepmaker09} and \citet{gouliermis15}, we fit the power law distribution to $1+\omega(\theta)$, similar as well to the work of \citet{odekon08}. For values of $\omega(\theta)>>1$, the power law fit will be the same regardless for the two methods. The distribution of star clusters in the galaxy disk is not homogeneous and their density depends on the galactocentric radius (see Section~\ref{sec:radial}). We do not have adequate numbers to divide our star cluster population by radius and age, and as such, all of our plots show the average correlation function between star cluster pairs averaged over the entire galaxy.

The physical motivation for applying the power law to $1+\omega(\theta)$ is that in a fully hierarchical (fractal) model, the distribution will have a smooth power law decline of $1 + \omega(\theta) \propto \omega^{\alpha}$ \citep{calzetti89} with increasing separation between pairs until they reach the correlation length of the hierarchy, where the clustering becomes consistent with being random ($1+\omega(\theta) = 1$). Inside such a distribution, the number of star clusters inside an annulus increases with the radius $r$ as $N=r^{\alpha} \times r^{2} \propto r^{\alpha + 2}$, relating the slope $\alpha$ measured from $1+\omega(\theta)$ directly to the two-dimension fractal number of the hierarchy as D2 $= \alpha + 2$, the fractal dimension for a distribution of objects that lie in the plane of a galaxy. We might expect the slope of $1+\omega(\theta)$ to change on scales smaller than the thickness of the disk but we do not have enough clusters to see the effect. Interstellar gas has a hierarchical morphology structure with a typical fractal dimension of $D2 \sim 1.5$ \citep{elmegreen06, sanchez08}. 
It is important to note that the fractal number does not uniquely describe a structure and it is difficult to distinguish true multi-scale fractal sub-clustering from the slope of the correlation function alone \citep{cartwright04, gouliermis14, lomax18}. However, for a scale-free distribution, the correlation function will have the form of a power law \citep{odekon08}. The correlation function thus allows us to constrain the difference in the clustering present between samples, where steeper slopes are consistent with expectations of fractal substructuring.  

\begin{figure}
\includegraphics[scale=0.52]{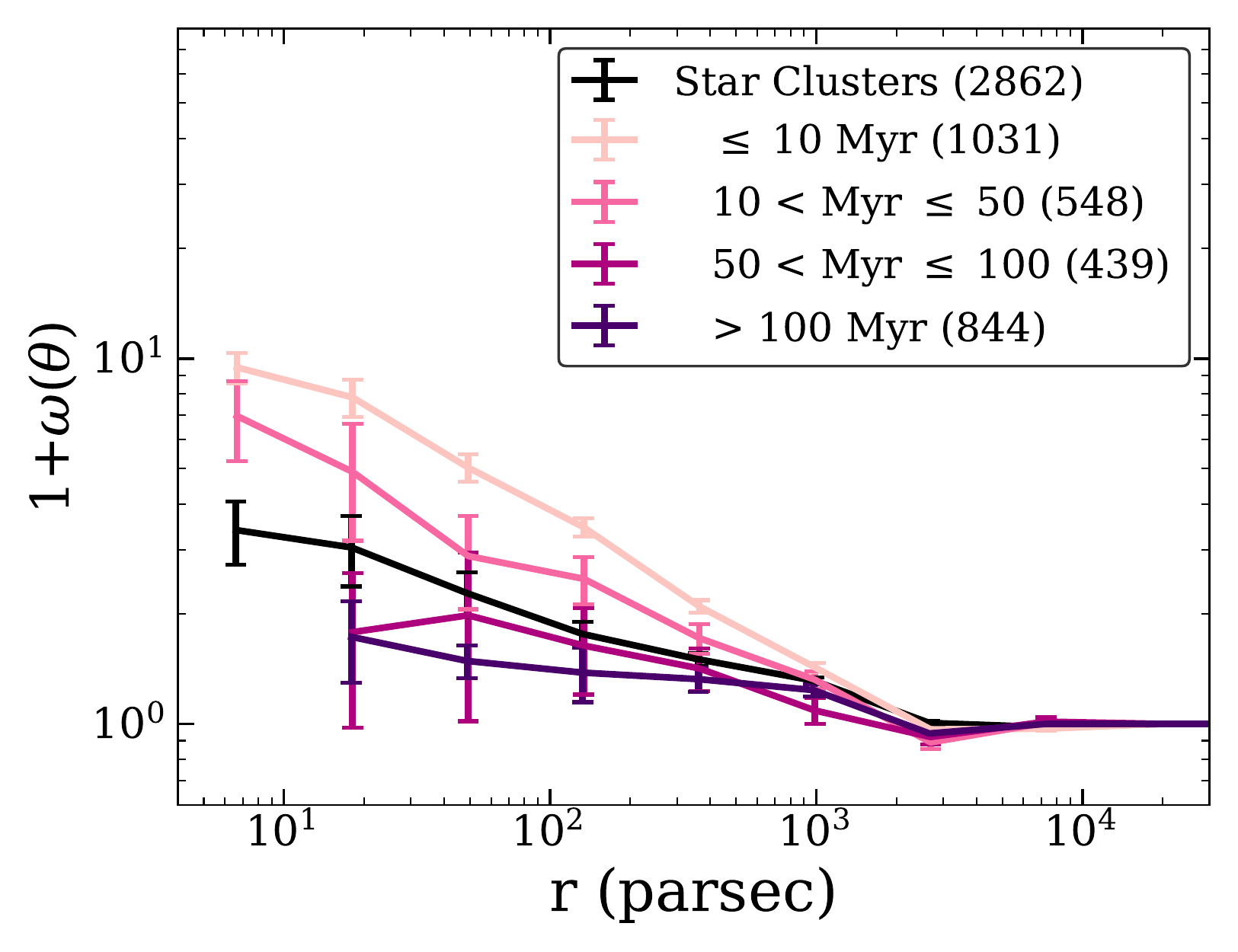}
\caption{
Two-point correlation function $1+\omega(\theta)$ for the star clusters across the entire disk of NGC~5194 (black line) as a function of physical scale. The star clusters are additionally divided into age ranges (pink lines; Section~\ref{sec:agecuts}). The numbers in parentheses show the number of clusters in each age group. There is a strong age dependency to the clustering, becoming consistent with a randomized distribution after 100~Myr. Additionally, there is a lack of clusters older than 50~Myr at small separation lengths compared to what is seen for younger clusters, further supporting that older clusters are less correlated with each other. 
\label{fig:2pcf}}
\end{figure}

Figure~\ref{fig:2pcf} shows the two-point correlation function for the star clusters across the disk of NGC~5194, where we exclude the 127 clusters in NGC~5195 (clusters above y-pixel values of 11500 in Figure~\ref{fig:5194}), reducing our total catalogue from 2989 to 2862 star clusters. We find a smooth and steady decline with increasing radius, well described with a power law. Younger star clusters show spatial distributions that are distinctly different than older star clusters, where the younger star clusters are more clustered than the older. The correlation function of star clusters older than $\sim$50--100~Myr, being almost flat, demonstrates that these clusters are more distributed than their younger counter parts. This behaviour is consistent with a hierarchical distribution and has been observed in the distribution of star formation, star clusters, and stars within other galaxies \citep{gomez93, zhang01, odekon06, odekon08, scheepmaker09, gouliermis14, gouliermis15, ali17} as well as the star clusters in other LEGUS galaxies \citep{grasha15, grasha17a}. The 1$\sigma$ errors for the correlation functions are bootstrap estimates based on 1000 samples and are not Poisson errors due to the correlation between the the errors; Poissonian approximations of the errors will underestimate the uncertainties in the parameters.
%

\subsubsection{Age Effects}\label{sec:agecuts}
The star clusters in NGC~5194 are preferentially associated with the spiral arms of the galaxy (Figure~\ref{fig:5194}). As shown in \citet{messa18a}, the concentration within the spiral arms is particularly noticeable for the very young clusters ($<$10~Myr), though the clustering persists for clusters up to ages of $\sim$100~Myr before diminishing. Thus, the hierarchical distribution slowly dissipates with time as the result of star clusters randomizing throughout the galaxy, creating a decline in $1 + \omega(\theta)$ with increasing age in addition to length scale. 

Figure~\ref{fig:2pcf} shows the correlation functions for clusters of different age ranges. We see a significant decrease in the slope and the amplitude of the clustering with an increase in the age of the clusters. The youngest clusters ($<$10~Myr) exhibit a slope of $-0.40\pm0.05$, significantly different from the global measurement of $-0.21\pm 0.03$. 50--100~Myr after their formation, the recovered slope is nearly flat at $-0.12 \pm 0.04$; the star clusters that are able to travel from their star-forming complexes take on distributions that are consistent with being randomized. Thus, clusters become less correlated with each other and the coherence of star formation as traced with star clusters subsides after 50--100~Myr; this randomization timescale is consistent with what we find in other local galaxies \citep{grasha17a}. While the randomization timescale is as short as 50~Myr, structures with lower fractal dimensions still survive for longer periods, as e.g., is found for stars in M31, where structure survives for at least 300~Myr \citep{gouliermis15a}, in accordance to what we found in NGC~6503 \citep{gouliermis15}.
%
\begin{table}
\caption{Power-Law Parameters of the two point correlation function. Columns list the 
(1) Subset of star clusters or GMCs;
(2) Number of star clusters or GMCs in each subset; 
(3) Amplitude $A$ of the angular correlation function; and
(4) Slope $\alpha$ of the angular correlation function. Numbers in parentheses indicate uncertainties in the final digit(s) of the listed quantities, when available.} 
\label{tab2}
 \begin{tabular}{lccc}
  \hline
Class				& 
N$_{\rm SC}$  	&
$A_{\omega}$	&  
$\alpha$  \\
  \hline
\multicolumn{4}{c}{All Star Clusters in NGC~5194}\\   
\hline
Star Clusters 									& 2862	& 5.4(0.5)		& $-0.21(0.03)$ \\ 
\hspace{5pt} $\leq$10 Myr 				& 1031	& 23(3)			& $-0.40(0.05)$ \\ 
\hspace{5pt} 10 $<$ Myr $\leq$ 50 	& 548	& 13(2)			& $-0.34(0.05)$ \\ 
\hspace{5pt} 50 $<$ Myr $\leq$ 100	& 439	& 2.7(0.2)		& $-0.12(0.04)$ \\ 
\hspace{5pt} $>$100 Myr 					& 844	& 2.1(0.2)		& $-0.07(0.03)$ \\
\hspace{5pt} $\leq$ 4 kpc 				& 1308	& 4.5(0.8)		& $-0.23(0.04)$ \\
\hspace{5pt} $>$ 4 kpc 					& 1554	& 14.9(0.9)	& $-0.32(0.02)$ \\
\hline\\
\multicolumn{4}{c}{Star Clusters in PAWS field$^b$}\\
\hline
Star Clusters 																						& 1268	& 4.1(0.8)	& $-0.19(0.05)$ \\
\hspace{5pt} $\leq$10 Myr 																	& 536	& 8(1) 		& $-0.28(0.04)$ \\
\hspace{5pt} $\leq$10 Myr \& Mass$>$ 5$\times10^{3}$ M$_{\odot}$	& 330 	& 10(1)		& $-0.31(0.04)$ \\
\hspace{5pt} $\leq$10 Myr \& Mass$>$ 3$\times10^{4}$ M$_{\odot}$	& 72	  	& 43(15)	& $-0.46(0.13)$ \\
\hline\\
\multicolumn{4}{c}{GMCs$^b$}\\
\hline
GMCs 																			& 1507	& 2.3(0.2)	& $-0.09(0.03)$ \\
\hspace{5pt} Mass$>$ 5$\times 10^{5}$ M$_{\odot}$ 	& 1070	& 2.5(0.3)	& $-0.11(0.04)$ \\
\hspace{5pt} Mass$>$ 3$\times 10^{6}$ M$_{\odot}$ 	& 338	& 8.4(1.0)	& $-0.27(0.04)$ \\
\hspace{5pt} Mass$>$ 5$\times 10^{6}$ M$_{\odot}$ 	& 169	& 20(3)		& $-0.35(0.05)$ \\
  \hline
  \multicolumn{4}{l}{$^a$ Excludes the star clusters in NGC~5195}\\
  \multicolumn{4}{l}{$^b$ The star clusters and the GMCs in the PAWS field are fit over the}\\
  \multicolumn{4}{l}{range 100--3000 pc}\\ 
 \end{tabular}
\end{table}

\subsubsection{Comparing the Clustering of Star Clusters to Molecular Clouds}\label{sec:2pcfpaws}
\begin{figure*}
\includegraphics[scale=0.49]{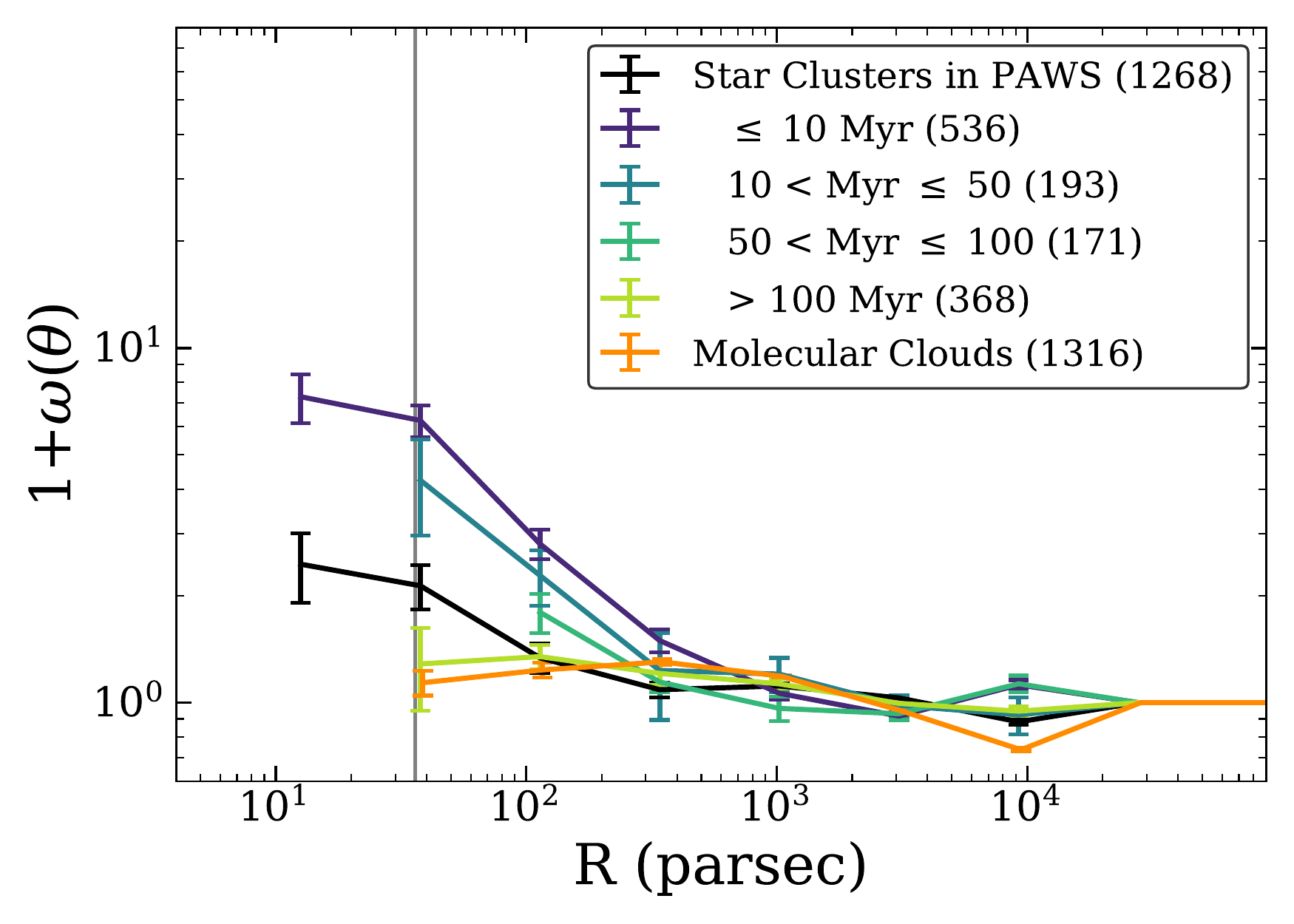}
\includegraphics[scale=0.49]{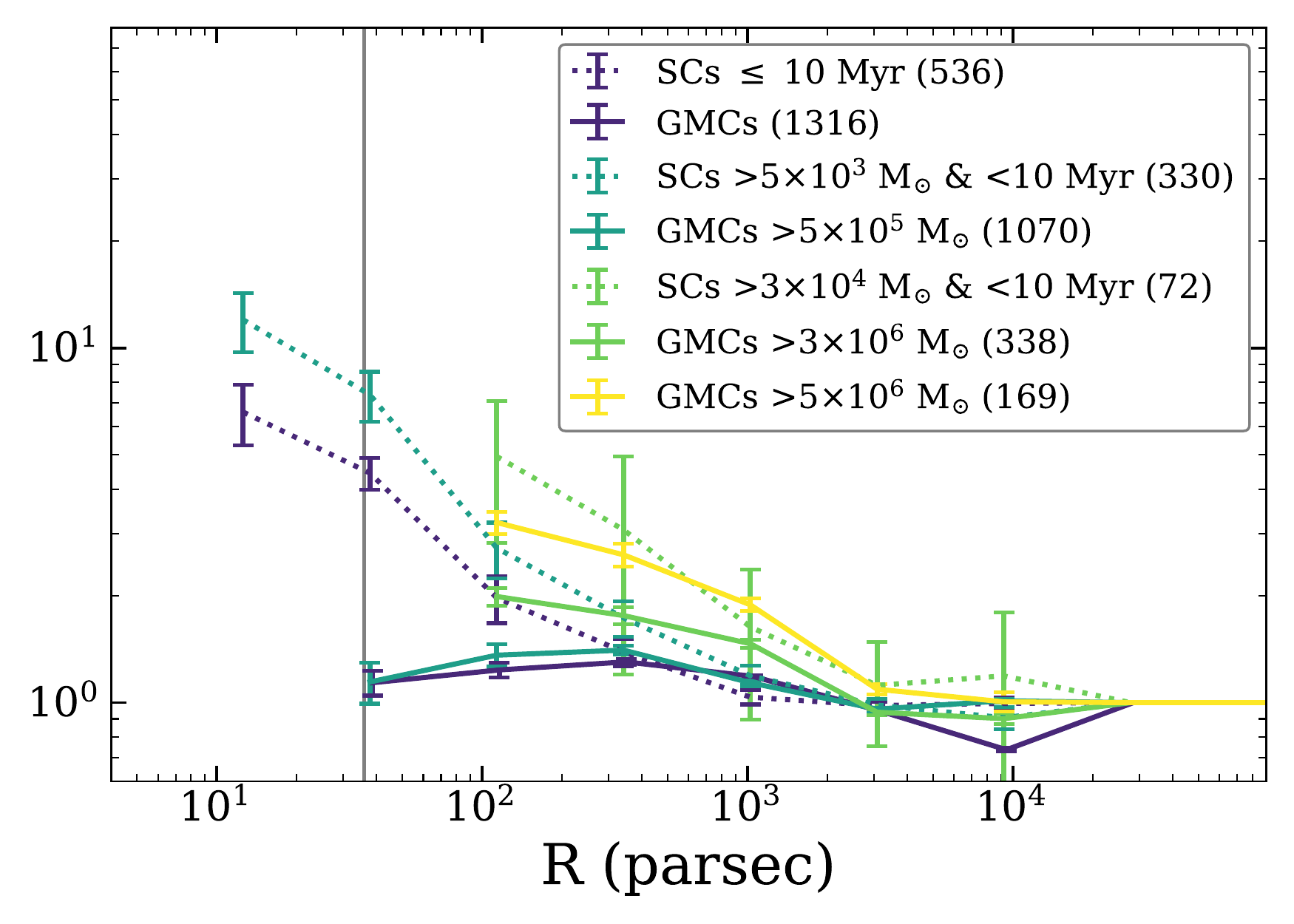}
\caption{
{\bf Left:} Two-point correlation function $1+\omega(\theta)$ for the GMCs and the clusters as a function of the spatial scale (parsec) that are located within both the UVIS and PAWS footprints. The numbers in parentheses show the number of clusters in each classification and the vertical dashed lines show the median radius of the GMCs at 36~pc. The GMCs have minimal structure, reaching a flat distribution accompanied by a slight excess at $\sim$800~pc. We separate the star clusters present in the PAWS coverage by age, finding that the youngest ($<$10~Myr) star clusters do not mirror the observed slope of the GMCs and the clustering is quickly lost around 50~Myr. The star clusters within the PAWS field show shallower slopes and smaller correlation lengths (see also Figure~\ref{fig:2pcfradial}) compared to the entire star cluster population in Figure~\ref{fig:2pcf}. 
{\bf Right:} A comparison between the distribution of the most massive and youngest star clusters to the most massive GMCs. We show star clusters (dashed purple line) with mass cuts at 5000~M$_{\odot}$ (dashed blue line) and $3\times10^4$~M$_{\odot}$ (dashed green line). We also show the equivalent GMC distribution (solid lines) assuming a 1\% SFE. The clustering in the distribution of the GMCs increase with mass and starts to resemble the distribution present in the star clusters, but at a given mass cutoff, the slopes of the GMCs are still different and significantly shallower from that observed for the youngest star clusters. The observed slope for the GMC distribution stars to increase for masses greater than $5\times10^6$~M$_{\odot}$ (yellow line), but there are not enough clusters to compare this distribution.  
\label{fig:2pcfpaws}}
\end{figure*}

Under the assumption that the young star clusters inherit their initial distribution from their natal clouds, we expect the initial stellar configurations to reflect the highly structured nature of the ISM gas. We again use the PAWS GMC catalogue \citep{colombo14a} to calculate the clustering present in the distribution of the molecular clouds within the inner 9 kpc. Figure~\ref{fig:2pcfpaws} shows the resulting correlation function for all the molecular clouds. The most noticeable difference between the clustering distribution of the star clusters and that of the molecular clouds is how the GMCs are located in a much more spatially homogeneous distribution compared to highly clustered distributions of the star clusters, a similar result found in simulations by \citet{parker15}. 

The second largest difference between the distribution of the star clusters and the GMCs in Figure~\ref{fig:2pcfpaws} compared to the global sample in Figure~\ref{fig:2pcf} is the correlation length where the function becomes consistent with a random distribution. The star clusters within the inner 9~kpc have correlation lengths ($1+\omega(\theta)=1$) of a few hundred parsec, significantly smaller than what is observed for the entire sample (Figure~\ref{fig:2pcfpaws}). Beyond this, the correlation signal is consistent with being random. The GMCs show a fairly large correlation length of $\sim$5000~pc, but part of this is due to the shallowness present in their distribution.
%

The measured slope for the GMCs (Table~\ref{tab2}) from the correlation function is $-0.10\pm 0.03$, significantly shallower than the subset of star clusters located within the PAWS region at $-0.28\pm0.04$, where we only measure the slope for the GMCs and the star clusters at distances between 100~pc and 3000~pc, as these scale lengths are covered fully by all our subsets. The difference in the measured slopes between star clusters and GMCs additionally increases when we only consider the youngest ages, where the hierarchy becomes substantially more clustered for clusters with ages less than 10~Myr old (Figure~\ref{fig:2pcfpaws}). The clustering of the GMCs is exceptionally flat compared to that of the star clusters, although it is consistent with that of star clusters older than 100~Myr. There is an excess in the clustering of the GMCs around a few hundred parsec, arising due to substructure present in the galaxy. While we expect the hierarchy of the youngest star clusters to mirror that of the GMCs, we find that this is not the case and that the slopes greatly deviate for the youngest clusters. 

The same trend for an observed shallower GMC distribution compared to the young star clusters is also observed within the flocculent galaxy NGC~7793 \citep{grasha18}. The results from these two galaxies illustrate that the excess in the distribution of the star clusters must arise from an inherently more clustered distribution compared to that of GMCs, indicating that not all GMCs result in a star cluster, and those that do must produce more than one star cluster, where the production of star clusters is sequential and not simultaneous. This comparison may be further hindered by our lack of sensitivity to the dense peaks of CO-dark molecular gas, making it difficult to detect the location of the dense ISM where the vast majority of H$_2$ may be actively forming \citep[e.g.,][]{grenier05, wolfire10, glover16}. 

To investigate this further, we compare the distribution of the youngest and most massive star clusters to the most massive GMCs (right panel of Figure~\ref{fig:2pcfpaws}). In attempt to match the mass limit between that of star clusters and GMCs, for a given cut in the star cluster mass, we assume a star cluster formation efficiency (SFE) of 1\% and use that to estimate the resulting stellar mass for a given GMC mass limit. SFE is expected to be of order a few percent in nearby galaxies \citep{leroy08, usero15}. There is a general increase in the slope of the GMC distribution with increasing mass, signifying that when we include all the molecular clouds in the correlation function, there is a washing out of the clustered signal as observed for stellar products. The most massive molecular clouds ($>$5$\times10^6$~M$_{\odot}$) start to show a significantly more clustered distribution, albeit with a slightly different distribution than that of the star clusters. However, we have inadequate numbers of young, massive star clusters above that value (star cluster mass $>$5$\times10^4$~M$_{\odot}$ for an assumed SFE of 1\%); we can only do the comparison for GMCs with a masscut at $>$3$\times10^6$~M$_{\odot}$ (star cluster mass $>$3$\times10^4$~M$_{\odot}$ for an assumed SFE of 1\%). We should note, however, that young star clusters are always significantly more clustered than massive GMCs (right panel of Figure~\ref{fig:2pcfpaws}).

The increased clustering for increasing GMCs mass implies that our optically identified star clusters arise from a specific subset of molecular clouds, partly explaining the difference in the spatial distribution in Figure~\ref{fig:2pcfpaws}. We conclude that it is the most massive clouds that are more likely to produce star clusters, given the comparison in their spatial distributions, although an improvement can be made in further studies with an increase in resolution of the CO gas to examine the relations at smaller spatial scales below our current resolution as well as the inclusion of dense gas tracers \citep[e.g.,][]{watanabe16}.

\subsubsection{Radial Trends}\label{sec:radial}
\begin{figure}
\includegraphics[scale=0.55]{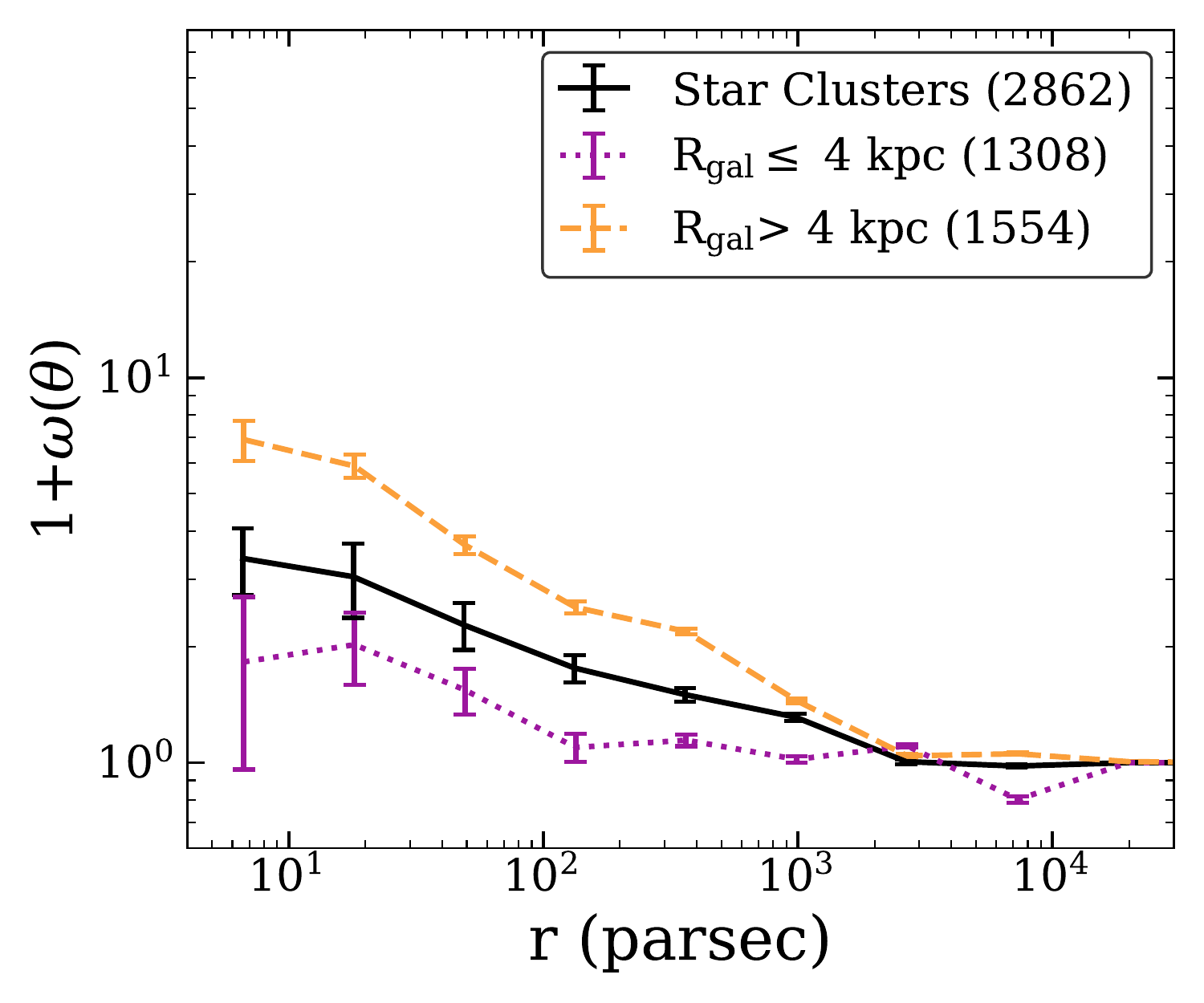}
\caption{
Two-point correlation function $1+\omega(\theta)$ for star clusters at galactocentric radii greater (orange dash line) or less (purple dot line) than 4~kpc. The numbers in parentheses show the number of clusters in each classification. Star clusters further from the center of the galaxy reside in larger hierarchies, exhibiting greater and slightly less shallow, correlation lengths than star clusters near the center. 
\label{fig:2pcfradial}}
\end{figure}

To investigate a potential radial dependence of the star cluster hierarchies, we divide the star clusters of NGC~5194 into two galactocentric bins, separated by the corotation radius of 4~kpc \citep{querejeta16}, and computing the correlation function for the clusters in these two radial bins. Figure~\ref{fig:2pcfradial} shows that the clustering does indeed depend on the position of the clusters within the stellar disk. For a given spatial scale, the clustering amplitude is significantly stronger for clusters located at distances greater than the co-rotation radius than for clusters within the co-rotation radius. More notably, the correlation length is also significantly larger, around $\sim$2~kpc, for clusters outside the co-rotation radius whereas the size of the structures for the clusters within the co-rotation radius is $\sim$200~pc. One of the biggest drivers of the observed difference is the lack of strong stellar spiral features within the inner region of the galaxy, resulting in significantly smaller star-forming complexes (Section \ref{sec:2pcfpaws} and Table~\ref{tab2}). 

This suggests longer lived star-forming regions reside at larger galactocentric radii. This is likely a result of lower shear within the outer regions of the galaxy, allowing clusters to live longer than in regions near the center of the galaxy. The effect of the local ambient environment in also driving the maximum correlation size is further discussed in Section \ref{sec:discussion}. 
%

\section{Discussion}\label{sec:discussion}
A comparable spatial distribution between the star clusters and molecular gas in the two-point correlation function would suggest an inheritance of the hierarchy between the components. Indeed, as shown in Section~\ref{sec:SCs_insideGMCs} and Figure~\ref{fig:dist_agemass}, star clusters that are still associated with molecular gas are ten times younger than star clusters that are spatially separated from molecular gas. As shown in Figure~\ref{fig:2pcfpaws}, despite the close association in time between GMCs and star formation, the hierarchy of the entire GMC catalog is very shallow compared to that of the star clusters. Not all GMCs are actively star-forming \citep{kawamura09} and we find that the distributions between the star clusters and the molecular gas become more similar under the assumption of a star formation efficiency of only a few percent. This mismatch between the star clusters and molecular gas was also demonstrated in the galaxy NGC~7793 in \citet{grasha18}. While requiring a small SFE identifies the molecular clouds that are more likely currently forming young star clusters, multiple effects are probably at play to result in the difference observed in the distributions. Turbulence may play a roll in modifying the shape of the correlation function \citep{federrath09}. Feedback effects are also important and quite possibly influence the distributions of star clusters differently from that of CO clouds, erasing any spatial correlation between the star clusters and the molecular gas on short timescales. The importance of feedback and turbulence at different length and time scales on the clustered distributions will be investigated in detail in future work. 

The local environment and location within the galaxy also impacts sizes of star-forming complexes. The correlation for star cluster complexes within the central 9~kpc exhibit correlation lengths of a few hundred parsec, significantly shortrer than the clustering scale length of 2~kpc for clusters located beyond 4~kpc (Figure~\ref{fig:2pcfradial}). By $\sim$50--100~Myr, the clustering signal has dispersed and is statistically the same as a random distribution, requiring a velocity of only $\sim$1~\kms\ to achieve randomization within this timescale. The shorter correlation sizes near the center are likely a result of the ambient environment that can influence the dissolution of individual clusters and unbound complexes through shear and pressure. The clustering becomes consistent with random at much smaller sizes in the inner region of the galaxy. This is consistent with the observed radial trend of clusters in the outer regions of NGC~5194 experiencing significantly less disruption \citep[e.g.,][]{silvavilla14, messa18b}. The decrease of shear, turbulence, and ISM mid-plane pressure with increasing distance from the galaxy center also increases the survivability of bound clusters, and hence, the lifetimes and sizes of the complexes in which they reside \citep[e.g.,][]{kruijssen12, johnson16}. Indeed, simulations do show that star and star cluster formation depends on the shear content in pre-collapse GMCs and that presence of shear in disk galaxies impedes the formation of very massive clusters, preferentially forming systems of smaller clusters and structures \citep{weidner10}. We would expect that if clusters cannot form, neither can clustering persist. While there is an age dependency on the amplitude of the clustering, younger star-forming regions do not necessarily exhibit shorter correlation lengths than older clusters (Figure~\ref{fig:2pcf}). A recent study in NGC~2336 found that UV-bright star-forming knots also show evidence for older star-forming regions at larger radii compared to the central part of the galaxy, quite possibly arising from the lower tidal shear present at larger galactocentric distances \citep{rahna18}. 

In \citet{grasha17b}, we find that the global average of young ($<$300~Myr) star-forming regions in NGC~5194 is $\sim$950~pc in size with average ages of 83~Myr, similar to the size scale recovered with the two-point correlation function in this work. From the parameters derived for a typical star-forming region of NGC~5194 from \citeauthor{grasha17b}, the velocity for a single crossing time is 13~\kms. This is consistent with the estimated velocity of 9.5~\kms\ required for star clusters to disassociate from their nearest molecular cloud given the distance and their current age from the closest GMC (Section \ref{sec:SCs_insideGMCs}). A study by \citet{whitmore14} in the Antennae galaxy shows that the timescale needed for GMC complexes with a radius of 200~pc to diffuse within a crossing time is $\sim$10~Myr. The results from Whitmore et al. are comparable to the values we derived in this work, with proportionally larger complex sizes and diffusion timescales within NGC~5194. 

The typical crossing timescale for spiral arm GMCs from PAWS, with a radius of 40~parsec and a velocity dispersion of 8~\kms\ is 5~Myr \citep{schinnerer17}. This is comparable to the 4--6~Myr timescale we derived for star clusters are still associated with their progenitor clouds. This suggests that the internal velocity, and thus, the crossing time scale of molecular clouds may determine how long a star cluster remains associated. The natural time unit for a GMC is the free-fall time and describes the timescale for which the cloud experiences no other forces outside of its own gravitational collapse. The free-fall timescale only depends on the density $\rho$ of the molecular gas, $t_{ff} = \sqrt{3\pi/32G \rho}$ and places a lower limit on a given cloud lifetime \citep{heyer15}. The lifetime of a typical GMC is expected to be a few free-fall timescales; observations show GMC lifetimes of $\sim$20--30~Myr within NGC~5194 \citep{meidt15}. The short timescales for clusters to become visible from their GMCs compared with the longer time-scales for which the clouds are destroyed is also expected from simulations \citep[e.g.,][]{dale15}. The $\sim$6~Myr timescale for the disassociation of star clusters from molecular gas will thus arise from both the relative time at which the young star clusters emerge due to secular motions as well as a dependency on the timescale for star clusters to erode cavities within their molecular reservoirs from feedback effects (i.e., ionizing winds). Star clusters are small in comparison to the angular extent of the GMCs, and GMC `destruction' remains a local process that does not encompass the GMC as a whole \citep{ochsendorf16}. 

An increase in the midplane pressure in the disk of a galaxy, corresponding to a higher surface density, can act to constrain the winds and movement of the star clusters, resulting in a longer timescale of association between the star clusters and gas. In \citet{grasha18}, we combine the star cluster catalogue of the flocculent galaxy NGC~7793 with $\sim$15~pc CO resolution from ALMA observations. Within NGC~7793, \citeauthor{grasha18} finds a shorter timescale of association between star clusters and GMCs of 2--3~Myr compared to the $\sim$6~Myr timescale we recover for NGC~5194 in this work. We conclude that the longer disassociation timescale in NGC~5194 is a result of an increase in the midplane-pressure in the disk compared to a lower surface density galactic system like NGC~7793. In agreement with our results, hydrodynamic simulations by \citet{kim18} show that cloud destruction takes $\sim$2--10~Myr after the onset of massive stellar feedback, with the disassociation timescale increasing with the gas surface density. 

It is clear that the onset of star formation and the dispersal of the immediate molecular material from the stellar radiative feedback occurs rapidly, on timescales less than 10~Myr \citep{clark05, hollyhead15, corbelli17, kim18, matthews18, miura18}, where our results show that the environment can play a role in the timescale for clusters to clear away their gas. Despite the stark differences between these two galactic systems, NGC~5194 in this study and NGC~7793 in \citet{grasha18}, the relative differences in the age of the star clusters as a function of distance from their nearest GMC remains the same; by the time a star cluster is within 2 radii of its nearest GMC, the star clusters will be half the age of the global star cluster population \citep[Figure~7 of][]{grasha18}, a result of the increase in the surface density raising the timescale that star clusters remain associated with their molecular gas. 

In a study of the Antennae galaxy, \citet{matthews18} finds that by 5~Myr, star clusters will start to lose association with their molecular gas and by 30~Myr, the majority of star clusters have lost all association with their molecular material. The Antennae galaxy is a similar system to NGC~5194 and thus it is expected for these systems to exhibit similar, albeit higher, timescales compared to what is seen in NGC~7793 \citep{grasha18}. Molecular clouds are observed to evolve over time and exhibit different levels of star formation activity. In the LMC, not all GMCs show evidence of star formation; more evolved and older clouds are more likely to be associated with optical stellar clusters \citep{kawamura09}. The same study derives a disassociation timescale between star clusters and GMCs of $\sim7-10$~Myr, however, no significant difference is observed in the properties (size or line width) between the varying evolutionary stages of the GMCs. Within this proposed picture for the evolutionary sequence of GMCs, most active star formation will lead to the dispersal of the cloud. The cloud population of NGC~5194 appears to undergo regrowth and transformation rather than a complete dispersal after star formation \citep{meidt15}, which suggests that the observed spatial separation between older star clusters and GMCs cannot arise solely from GMC dissolution and must require a component of secular motion that separates star clusters from their natal clouds.

\section{Summary and Conclusion}\label{sec:summary}
The LEGUS project \citep{calzetti15a} has produced an extensive suite of high-quality, homogeneous, and accurate properties of clusters in nearby galaxies, which are crucial in order to address the role of star clusters in the star formation process in a consistent manner across galaxies. In this paper, we combine the LEGUS star cluster catalogue in NGC~5194 with the GMC catalogue from the PAWS survey \citep{schinnerer13} to study the timescale of association between the star clusters and the molecular gas. 

Our main results can be summarized as follows:
\begin{enumerate}
	\item The locations of the youngest star clusters are highly correlated with the GMC catalogue from PAWS data. Star clusters that are associated (i.e., located within the footprint of a GMC) are 10 times younger with median ages of 4~Myr than star clusters that are spatially unassociated with molecular clouds, which exhibit median ages of 50~Myr. 
	
	\item Clusters that are at least as old as the typical lifetime of molecular clouds ($\gtrsim$20~Myr) or older are located at distances of at least three radii from their nearest GMC. Thus, the timescale we derive for the disassociation of star clusters with molecular clouds ($\sim$4--6~Myr) is impacted by both the overall dissolution of the GMC and the secular motions that move the clusters away from their initial birth locations. We also find no significant evidence that more massive GMCs are more likely to host multiple star clusters (Figure~\ref{fig:massGMCvsnumSC}). 
	
	\item We derive the two-point correlation function to quantify the time-scale for the survival of star-forming hierarchical structures and to determine the sizes of the clustered regions, allowing us to constrain the typical lifetimes of $z\sim0$ star-forming regions. We find that the amount of clustering of star clusters decreases with increasing spatial scale (Figure~\ref{fig:2pcf}), consistent with star cluster formation occurring within hierarchical star-forming complexes of $\sim$1~kpc in size. The power law slope of the distribution of the star clusters across the galactic disk is $\alpha = -0.21 \pm 0.03$. The clustering depends strongly on the age of the clusters, with the clusters younger than 10~Myr showing a slope of $\alpha = -0.40 \pm 0.05$, indicative that star-forming hierarchies disperse with time, with randomization timescales of $\sim$50--100~Myr within NGC~5194. 
	
	\item The correlation function for the GMCs shows a power law slope of $-0.09 \pm 0.03$, very shallow compared to the star clusters, consistent with being nearly randomized (Figure~\ref{fig:2pcfpaws}). The GMCs are distributed in a more homogeneous spatial manner than the star clusters; this may suggest that the large-scale structure of GMCs is more easily randomized compared to that of star clusters. When we match the mass limit of the star clusters to that of the GMCs by assuming a star formation efficiency of a few percent, we find that the clustering present in the GMCs increases to $-0.35\pm0.05$ for the most massive GMCs ($>$5$\times10^6$~\Msol). This allows for a better identification of the subset of GMCs that are forming the current population of star clusters and that the clustering present in the youngest and most massive star clusters better reflects the clustering present in most massive clouds. 
	
	\item The size and the strength of the clustering also depend on the galactocentric radius; larger clustered star-forming regions are preferentially located further from the galaxy center (Figure~\ref{fig:2pcfradial}). This environmental dependence of the clustering of the star clusters is consistent with clusters near the center experiencing increased disruption in a region with higher turbulence, midplane pressure, and shear. This is in agreement with the strong age-dependency of the survival of the cluster complexes with complexes near the inner region of the galaxy being smaller and showing distributions that are close to random. 
	
\end{enumerate}

\section*{Acknowledgements}
We are grateful to an anonymous referee for careful reading of this manuscript and for providing comments that improved the scientific outcome and quality of the paper.
We appreciate the enlightening discussions and valuable comments on this work by D.A. Hunter and A.J. Battisti. Based on observations made with the NASA/ESA Hubble Space Telescope, obtained at the Space Telescope Science Institute, which is operated by the Association of Universities for Research in Astronomy, under NASA Contract NAS 5--26555. These observations are associated with Program 13364 (LEGUS). Support for Program 13364 was provided by NASA through a grant from the Space Telescope Science Institute. 
This research has made use of the NASA/IPAC Extragalactic Database (NED) which is operated by the Jet Propulsion Laboratory, California Institute of Technology, under contract with NASA. 
Parts of this research were supported by the Australian Research Council Centre of Excellence for All Sky Astrophysics in 3 Dimensions (ASTRO 3D), through project number CE170100013.
AA acknowledges the support of the Swedish Research Council (Vetenskapsr\r{a}det) and the Swedish National Space Board (SNSB). DAG kindly acknowledges financial support by the German Research Foundation (DFG) through programme GO 1659/3-2. 
KG acknowledges the supportive and collaborative environment provided by the ASTRO 3D writing retreat at Batemans Bay.

\textit{Software}:
Astropy \citep{astropy}, 
iPython \citep{ipython}, 
Matplotlib \citep{matplotlib}, 
Numpy \citep{numpy}, 
TreeCorr \citep{treecorr}

\bibliographystyle{mnras}


\bsp	
\label{lastpage}
\end{document}